\documentclass[floats,floatfix,showpacs,amssymb,prd,twocolumn,superscriptaddress,nofootinbib]{revtex4-1}

\usepackage{graphicx,epsf, epsfig, amssymb}
\usepackage{bm}
\usepackage{longtable}
\usepackage[usenames,dvipsnames]{xcolor}
\usepackage{color}
\usepackage[breaklinks]{hyperref}
\usepackage{amsfonts,amsmath,amssymb,mathrsfs}
\usepackage{multirow}
\usepackage[nolist,nohyperlinks]{acronym}
\usepackage{xspace}
\usepackage{booktabs,array}
\usepackage[british]{babel}
\usepackage[utf8]{inputenc}

\newcommand{\tn}{\textnormal}

\begin{document}

\title{Slowly rotating black hole solutions in Horndeski gravity}

\author{Andrea Maselli}
\email{andrea.maselli@roma1.infn.it}
\affiliation{Center for Relativistic Astrophysics, School of Physics,
Georgia Institute of Technology, Atlanta, Georgia 30332, USA}

\author{Hector O. Silva}
\email{hosilva@phy.olemiss.edu}
\affiliation{Department of Physics and Astronomy, The University of Mississippi, University, Mississippi 38677, USA}

\author{Masato Minamitsuji}
\email{masato.minamitsuji@ist.utl.pt}
\affiliation{Departamento de F\'isica, CENTRA, Instituto Superior
T\'ecnico, Universidade de Lisboa, Avenida Rovisco Pais 1,
1049 Lisboa, Portugal}

\author{Emanuele Berti}
\email{eberti@olemiss.edu}
\affiliation{Department of Physics and Astronomy, The University of Mississippi, University, MS 38677, USA}
\affiliation{CENTRA, Departamento de F\'isica, Instituto Superior
T\'ecnico, Universidade de Lisboa, Avenida Rovisco Pais 1,
1049 Lisboa, Portugal}


\date{\today}

\begin{abstract}
  We study black hole solutions at first order in the Hartle-Thorne
  slow-rotation approximation in Horndeski gravity theories. We derive
  the equations of motion including also cases where the scalar
  depends linearly on time.  In the Hartle-Thorne formalism, all
  first-order rotational corrections are described by a single
  frame-dragging function. We show that the frame-dragging function is
  exactly the same as in general relativity for all known black hole solutions
  in shift symmetric Horndeski theories, with the exception of
  theories with a linear coupling to the Gauss-Bonnet invariant. Our
  results extend previous no-hair theorems for a broad class of
  Horndeski gravity theories.
\end{abstract}

\maketitle


\section{Introduction}
General relativity (GR) has passed all experimental tests in the Solar
System and in binary pulsars with flying
colors~\cite{Will:2014kxa}. Current observations mostly probe the
weak-field/slow-motion regime of the theory (with the exception of
binary pulsars, where the orbital motion is nonrelativistic but the
individual binary members are compact objects), and some of the most
interesting strong-field predictions of GR are still elusive and
difficult to verify.
Observational and theoretical issues with Einstein's theory --
including the dark matter and dark energy problems, the
origin of curvature singularities and the quest for an ultraviolet
completion of GR -- have motivated strong efforts to develop modified
theories of gravity which differ from GR in the infrared and
ultraviolet regimes, while being consistent with the stringent
observational bounds at intermediate energies~\cite{Berti:2015itd}.
The search for unambiguous signatures of modifications of GR in the
strong-gravity regime is a major goal of several research fields,
including cosmology~\cite{Clifton:2011jh}, ``standard''
electromagnetic astronomy~\cite{Psaltis:2008bb}, and Earth- and
space-based gravitational-wave
astronomy~\cite{Yunes:2013dva,Gair:2012nm}.

In this work we consider a class of modifications of GR known as
Horndeski gravity~\cite{Horndeski:1974wa}. This is the most general
scalar-tensor theory with a single scalar yielding second-order field
equations for the metric and the scalar field (see
e.g. \cite{Damour:1992we,Horbatsch:2015bua} for tensor-multiscalar
theories, and \cite{Padilla:2012dx,Charmousis:2014zaa} for multiscalar
versions of Horndeski gravity).
All the terms present in the action of Horndeski gravity have been
shown to be originating from Galileons, i.e. scalar-tensor models
having Galilean symmetry in flat
space-time~\cite{Nicolis:2008in}. ``Generalized Galileon'' theories in
curved space-time in any number of dimensions were studied
in~\cite{Deffayet:2009mn}, and shown to be equivalent to Horndeski
gravity in four dimensions in~\cite{Kobayashi:2011nu}.
Furthermore, Horndeski gravity can be shown to emerge from a
Kaluza-Klein compactification of higher-dimensional Lovelock gravity
(see e.g.~\cite{Charmousis:2014mia} for an introduction to this topic,
and for a discussion of the relation between exact solutions in
Lovelock and Horndeski gravity).

The equations of motion of Horndeski gravity can be derived from the
action
\begin{equation}\label{Haction}
S=\sum_{i=2}^{5}\int d^{4}x\sqrt{-g}{\cal L}_i\ ,
\end{equation}
where
\begin{eqnarray}
{\cal L}_2&=&G_2\ ,\\
{\cal L}_3&=&-G_{3}\square\phi\ ,\\
{\cal L}_4&=&G_{4}R+G_{4\tn{X}}\left[(\square\phi)^2-\phi_{\mu\nu}^2\right]\ ,\\
{\cal L}_5&=&G_{5}G_{\mu\nu}\phi^{\mu\nu}\nonumber\\
&&-\frac{G_{5\tn{X}}}{6}\left[(\square\phi)^3+2\phi_{\mu\nu}^3\
-3\phi_{\mu\nu}^2\square\phi\right]\ .
\end{eqnarray}
Here $g_{\mu\nu}$ is the metric tensor, $g\equiv {\rm det}(g_{\mu\nu})$,
and $R$ and $G_{\mu\nu}$ are the Ricci scalar and the Einstein tensor
associated with $g_{\mu\nu}$, respectively.
We have introduced the functions $G_{i}=G_{i}(\phi,X)$, which depend
only on the scalar field $\phi$ and its kinetic energy
$X=-\partial_\mu\phi\partial^\mu\phi/2$, and we use units such that the
reduced Planck mass $m_{\rm Pl}^2=(8\pi G)^{-1}=1$.
For brevity we have
also defined the shorthand notation
$\phi_{\mu\dots\nu}\equiv \nabla_\mu\dots\nabla_\nu\phi$,
$\phi_{\mu\nu}^2 \equiv \phi_{\mu\nu}\phi^{\mu\nu}$,
$\phi_{\mu\nu}^3 \equiv
\phi_{\mu\nu}\phi^{\nu\alpha}\phi^{\mu}{_{\alpha}}$
and $\Box\phi\equiv g^{\mu\nu} \phi_{\mu\nu}$.
Horndeski theories are an interesting phenomenological playground for
strong-field gravity because they include as special cases all dark
energy and modified gravity models with a single scalar degree of
freedom:
\begin{itemize}
\item[(1)] the GR limit corresponds to $G_4(\phi,X)=1/2$, with
  $G_2=G_3=G_5=0$;
\item[(2)] when the only nonzero term is $G_4(\phi,X)=F(\phi)$ we
  recover a scalar-tensor theory with nonminimal coupling of the form
  $F(\phi)R$, and therefore Brans-Dicke theory and $f(R)$ gravity are
  special cases of Horndeski gravity;
\item[(3)] Einstein-dilaton-Gauss-Bonnet (EdGB) gravity, i.e. a theory
  with action
  \begin{equation}\label{EdGBaction}
    S=\int d^4x
\sqrt{-g}\left(\frac{1}{2}R+X+\xi(\phi) R^2_{\tn{GB}}\right)\ ,
  \end{equation}
  where
  $R^2_\tn{GB}=R^2-4R_{\mu\nu}R^{\mu\nu}+R_{\alpha\beta\gamma\delta}R^{\alpha\beta\gamma\delta}$
  is the Gauss-Bonnet invariant, corresponds to setting
\begin{align}
G_2=&X+8\xi^{(4)} X^2(3-\ln X)\ ,\label{EDGBK}\\
G_3=&4\xi^{(3)}X(7-3\ln X)\ ,\\
G_4=&\frac{1}{2}+4\xi^{(2)}X(2-\ln X)\ ,\\
G_5=&-4\xi^{(1)}\ln X\ ,\label{EDGBG5}
\end{align}
where $R_{\alpha\beta\gamma\delta}$ and $R_{\mu\nu}$ are the Riemann and Ricci tensors,
and we have defined $\xi^{(n)}\equiv \partial^n \xi/\partial \phi^n$ \cite{Kobayashi:2011nu};
\item[(4)] a theory with nonminimal derivative coupling of the form
\begin{equation}\label{nonminact}
S=\int d^{4}
x\sqrt{-g}\left[\zeta R+2\eta X+\beta G^{\mu\nu}\phi_\mu\phi_\nu
-2\Lambda_0\right]
\end{equation}
(see
e.g.~\cite{Sushkov:2009hk,Saridakis:2010mf,Germani:2010gm,Germani:2010hd,Gubitosi:2011sg}
for cosmological studies of this type of action) corresponds to the
following choice\footnote{A coupling of the form
  $G^{\mu\nu}\phi_{\mu}\phi_{\nu}$ can also be obtained by setting
  $G_5=-\phi$ and integrating by parts.} of the coupling
functions~\cite{Kobayashi:2014eva}:
\begin{align}
G_{2}&=-2\Lambda_0+2\eta X\ , \\
G_4&=\zeta+\beta X\ , \\
G_3&=G_5=0\ ,
\end{align}
where $\Lambda_0$, $\eta$, $\zeta$ and $\beta$ are
constants;
\item[(5)] the Lagrangian ${\cal L}_2$ corresponds to the k-essence
  field \cite{ArmendarizPicon:2000ah,ArmendarizPicon:1999rj,Alishahiha:2004eh}
  (and therefore part of the literature uses a different
  notation, where $G_2$ is denoted by $K$);
\item[(6)] the covariant Galileon of Ref.~\cite{Deffayet:2009wt} is
  recovered by setting $G_2=-c_2X$, $G_3=-c_3 X/M^3$,
  $G_4=M_{\rm Pl}^2/2-c_4X^2/M^6$ and $G_5=3c_5X^2/M^9$, where the
  $c_i$ ($i=2,\ldots,5$) are constants and $M$ is a constant with dimensions of mass.
\end{itemize}

In this paper we are interested in black hole (BH) solutions in
Horndeski gravity. As one of the most striking strong-field
predictions of GR, BHs are ideal astrophysical laboratories to test
gravity in the strong-field regime.
Various authors explored nonrotating BH solutions in special classes
of Horndeski gravity. Rinaldi~\cite{Rinaldi:2012vy} studied BH
solutions in theories with a nonminimal derivative coupling to the
Einstein tensor of the form $G^{\mu\nu}\phi_{\mu} \phi_{\nu}$.
Minamitsuji~\cite{Minamitsuji:2013ura} and Anabalon {\it et
  al.}~\cite{Anabalon:2013oea} found more general solutions by adding
a cosmological constant.  Kobayashi and
Tanahashi~\cite{Kobayashi:2014eva} studied BH solutions in a subclass
of Horndeski theories that is both shift symmetric (i.e., symmetric
under $\phi\to \phi+c$, with $c$ a constant) and reflection symmetric
(i.e., symmetric under $\phi\to -\phi$).
Under these assumptions, the only nonzero terms in the action are
${\cal L}_2$ and ${\cal L}_4$.  Theories with nonminimal derivative
coupling are both shift and reflection symmetric, and therefore they
are a subclass of the theories considered in
Ref.~\cite{Kobayashi:2014eva}.

The nonrotating BH solutions found in the works listed above either
reduce to the Schwarzschild solution or are not asymptotically
flat. This is a consequence of the no hair theorem by Hui and
Nicolis~\cite{Hui:2012qt}, which states that vacuum, static,
spherically symmetric, asymptotically flat BHs have no-hair in
Horndeski theories {\em with shift symmetry}. As pointed out by
Sotiriou and Zhou~\cite{Sotiriou:2013qea,Sotiriou:2014pfa} the theorem
actually has a loophole: asymptotically flat solutions can exist for
theories of the EdGB type with $\xi(\phi)=\phi$ (these theories are
still shift symmetric, because the Gauss-Bonnet combination is a
topological invariant). The solutions found in
Refs.~\cite{Sotiriou:2013qea,Sotiriou:2014pfa} are effectively special
cases of the nonrotating EdGB BH solutions studied by Kanti {\it et
  al.}~\cite{Kanti:1995vq}, that were subsequently generalized to slow
rotation in Refs.~\cite{Pani:2009wy,Ayzenberg:2014aka,Maselli:2015tta}
and to rapid rotation in
Refs.~\cite{Kleihaus:2011tg,Kleihaus:2014lba}. Other possibilities to
violate the no-hair theorems include adding a time dependence to the
scalar (but not to the metric), as in the solution proposed by
Babichev and Charmousis~\cite{Babichev:2013cya}, or considering
biscalar extensions of Horndeski gravity~\cite{Charmousis:2014zaa}.
Reference~\cite{Babichev:2015rva} extended the solutions in
\cite{Babichev:2013cya} to the charged case, allowing for a coupling
of the derivative of the scalar field to the energy-momentum tensor of
the Maxwell field.  Reference~\cite{Kolyvaris:2011fk} argued that a phase
transition to charged hairy BHs can be realized through a nonminimal
derivative coupling to the Einstein tensor; in this case, however, the
equations of motion were solved perturbatively.

The key question we address in this paper is the following: does rotation produce
interesting violations of the no-hair theorem at leading order in a
slow-rotation expansion? In experimental terms, could we possibly
observe violations of the no-hair theorem via frame-dragging
experiments? The conclusion of our analysis is that frame-dragging
corrections are {\em exactly the same as in GR} for all of the
Horndeski BH solutions that we analyzed, with the (already known)
exception of BH solutions in EdGB
gravity~\cite{Kanti:1995vq,Pani:2009wy,Sotiriou:2013qea,Sotiriou:2014pfa,Ayzenberg:2014aka,Maselli:2015tta}. We
do not expect this result to hold at second order in rotation, and
this will be the topic of a follow-up study.

The plan of the paper is as follows. In Sec.~\ref{sec:EOM} we present
the equations of motion for slowly rotating BH space-times in
Horndeski gravity, and we carry out some basic sanity checks (in
particular, we check that GR and EdGB gravity are recovered in the
appropriate limits). The field equations themselves are rather
lengthy, and they are listed in Appendix~\ref{app:FE} for the reader's
convenience. In Sec.~\ref{sec:nonmin} we study slowly rotating BHs in
theories with a nonminimal derivative coupling with the Einstein
tensor, finding that frame-dragging corrections are {\em exactly the
  same as in GR}. In Sec.~\ref{sec:current} we provide arguments
(based on the work of
Refs.~\cite{Hui:2012qt,Sotiriou:2013qea,Sotiriou:2014pfa}) to support
this no-hair result. Finally, in Sec.~\ref{sec:concl} we present some
conclusions and point out directions for future work.


\section{The equations of motion}
\label{sec:EOM}

The equations of motion that follow from the action~\eqref{Haction}
can be written schematically as ${\cal E}_{\alpha\beta}=0$ (from
variations of the metric) and ${\cal E}_{\phi}=0$ (from variations of
the scalar field), where
\begin{widetext}
\begin{align}
{\cal E}_{\alpha\beta}=&-\frac{g_{\alpha\beta}}{2}G_2+G_{2\tn{X}}X_{\alpha\beta}-\left[G_{3\tn{X}}X_{\alpha\beta}\square\phi+\frac{1}{2}g_{\alpha\beta}G_{3\mu}\phi^{\mu}-G_{3(\alpha}\phi_{\beta)}\right]+{G}_{\alpha\beta}G_4+G_{4\tn{X}}X_{\alpha\beta}R\nonumber\\
&+G{_{4\mu}}^{\mu}g_{\alpha\beta}-G_{4\alpha\beta}+
\left[G_{4\tn{XX}}X_{\alpha\beta}-\frac{1}{2}G_{4\tn{X}}g_{\alpha\beta}\right](\square\phi^2-\phi_{\mu\nu}^2)+2\square\phi G_{4\tn{X}}\phi_{\alpha\beta}-2\nabla_{(\alpha}[G_{4\tn{X}}\phi_{\beta)}\square\phi]\nonumber\\
&+\nabla_{\mu}[G_{4\tn{X}}\phi^{\mu}\square\phi]g_{\alpha\beta}+
2\nabla_{\mu}[G_{4\tn{X}}\phi_{(\alpha}\phi^{\mu}{_{\beta)}}]-\nabla_{\mu}[G_{4\tn{X}}\phi^{\mu}\phi_{\alpha\beta}]
-2G_{4\tn{X}}\phi_{\beta\nu}\phi^{\nu}{_{\alpha}}+{G}_{\mu\nu}\phi^{\mu\nu}(G_{5\tn{X}}X_{\alpha\beta}\nonumber\\
&-\frac{1}{2}G_5g_{\alpha\beta})+2G_5\phi^{\mu}{_{(\beta}}{G}_{\alpha)\mu}-\nabla^{\mu}[G_{5}\phi_{(\alpha}{G}_{\beta)\mu}]+\frac{1}{2}\nabla^{\mu}[G_{5}\phi_{\mu}{G}_{\alpha\beta}]+\frac{1}{2}\big\{RG_5\phi_{\alpha\beta}-R_{\alpha\beta}G_5\phi{_{\mu}}^{\mu}\nonumber\\
&+\square(G_5\phi_{\alpha\beta})+\nabla_{\alpha}\nabla_{\beta}(G_5\phi{_{\mu}}^{\mu})-2\nabla_{\mu}\nabla_{(\alpha}[G_5\phi{_{\beta)}}^{\mu}]+g_{\alpha\beta}[\nabla_{\mu}\nabla_{\nu}(G_5\phi^{\mu\nu})-\square(G_5\phi{_{\nu}}^{\nu})]\big\}\nonumber\\
&-\frac{1}{6}(G_{5\tn{XX}}X_{\alpha\beta}-\frac{1}{2}g_{\alpha\beta}G_{5\tn{X}})[(\square\phi)^3+2\phi_{\mu\nu}^3
-3\phi_{\mu\nu}^2\square\phi]-\frac{1}{2}\big\{G_{5\tn{X}}(\square\phi)^2\phi_{\alpha\beta}-2\nabla_{(\alpha}[G_{5\tn{X}}(\square\phi)^2\phi_{\beta)}]\nonumber\\
&+\frac{1}{2}g_{\alpha\beta}\nabla^{\mu}[G_{5\tn{X}}(\square\phi)^2\phi_{\mu}]\big\}-\big\{G_{5\tn{X}}\phi_{\mu\alpha}\phi_{\beta\sigma}\phi^{\sigma\mu}-\nabla^{\sigma}[G_{5\tn{X}}\phi_{(\alpha}\phi_{\mu\sigma}\phi^{\mu}{_{\beta)}}]+
\frac{1}{2}\nabla^{\sigma}[G_{5\tn{X}}\phi_{\sigma}\phi_{\mu\alpha}\phi^{\mu}{_{\beta}}]\big\}\nonumber\\
&+\frac{1}{2}\big\{G_{5\tn{X}}(\phi_{\mu\nu}^2\phi_{\alpha\beta}+2\square\phi\phi_{\alpha\mu}\phi^{\mu}{_{\beta}})
-\nabla_{(\beta}[G_{5\tn{X}}\phi_{\alpha)}\phi_{\mu\sigma}\phi^{\mu\sigma}]
+\frac{1}{2}g_{\alpha\beta}\nabla^{\sigma}[G_{5\tn{X}}\phi_{\sigma}\phi_{\mu\nu}\phi^{\mu\nu}]\nonumber\\
&-2\nabla^{\mu}[G_{5\tn{X}}\square\phi\phi_{(\alpha}\phi_{\beta)\mu}]+\nabla^{\mu}[G_{5\tn{X}}\square\phi\phi_{\mu}\phi_{\alpha\beta}]\big\}\ ,
\label{eab}
\end{align}
\begin{align}
{\cal E}_{\phi}=&\phantom{+}G_{2\phi}+\nabla_{\alpha}(G_{2\tn{X}}\phi^{\alpha})-G{_{3\alpha}}^{\alpha}-\nabla_{\alpha}(G_{3\tn{X}}\phi^{\alpha}\square\phi)-\square\phi G_{3\phi}+G_{4\phi}R+(\square\phi^2-\phi_{\mu\nu}^2)G_{4\tn{X}\phi}\nonumber\\
&+\nabla^{\alpha}[G_{4\tn{XX}}\phi_{\alpha}(\square\phi^2-\phi_{\mu\nu}^2)]+\nabla^{\alpha}(G_{4\tn{X}}\phi_{\alpha}R)
+2\square(G_{4\tn{X}}\square\phi)-2\nabla^{\alpha}\nabla^{\beta}(G_{4\tn{X}}\phi_{\alpha\beta})+G_{5\phi}{G}_{\alpha\beta}
\phi^{\alpha\beta}\nonumber\\
&+G_{5}^{\alpha\beta}{G}_{\alpha\beta}-\frac{1}{6}G_{5\tn{X}\phi}[(\square\phi)^3+2\phi_{\mu\nu}^3
-3\phi_{\mu\nu}^2\square\phi]+\nabla^{\alpha}[G_{5\tn{X}}\phi_{\alpha}\phi_{\mu\nu}{G}^{\mu\nu}]-\frac{1}{6}
\nabla^{\alpha}\{G_{5\tn{XX}}\phi_{\alpha}[(\square\phi)^3\nonumber\\
&+2\phi_{\mu\nu}^3-3\phi_{\mu\nu}^2\square\phi]\}-\frac{1}{2}\square[G_{5\tn{X}}(\square\phi)^2]
-\nabla^{\alpha}\nabla^{\beta}[G_{5\tn{X}}\phi^{\mu}{_{\alpha}}\phi_{\mu\beta}]
+\frac{1}{2}\square(G_{5\tn{X}}\phi_{\mu\nu}^2)+\nabla^{\alpha}\nabla^{\beta}(G_{5\tn{X}}\phi_{\alpha\beta}\square\phi)\ .
\label{ep}
\end{align}
\end{widetext}
Here we have defined $G_{i\alpha}\equiv \nabla_{\alpha}G_{i}$,
$X_{\alpha\beta}\equiv \delta X/\delta g_{\alpha\beta}$, and
$f_{1(\alpha}f_{2\beta)}\equiv
(f_{1\alpha}f_{2\beta}+f_{1\beta}f_{2\alpha})/2$.
These equations apparently contain higher derivatives, but they can be
shown to be of second order using appropriate identities
(cf. e.g. Appendix B of~\cite{Kobayashi:2011nu}).

To investigate the properties of slowly rotating BH solutions in
Horndeski gravity we follow the approach developed by
Hartle~\cite{Hartle:1967he,Hartle:1968si}, in which rotational
corrections to the static, spherically symmetric background are
introduced within a perturbative framework.
At linear order in the BH angular velocity $\Omega$, the metric can be
written in the form
\begin{equation}\label{Hartle2}
ds^2=-A(r)dt^2+\frac{dr^2}{B(r)}+r^2(d\theta^2+\sin^2\theta
  d\varphi^2)
-2\omega(r)dtd\varphi,
\end{equation}
where the frame-dragging function $\omega(r)$ is of order $\Omega$.

Kobayashi {\it et al. }~\cite{Kobayashi:2012kh,Kobayashi:2014wsa}
carried out a fully relativistic analysis of linear perturbations
around static, nonrotating, spherically symmetric backgrounds. As a
preliminary step for this perturbative analysis, they derived the
equations of motion for general static, spherically symmetric vacuum
space-times. Here we generalize these results to the slowly rotating
case, deriving the equations of motion for the metric component
$\omega(r)$. We also generalize the analysis of
Refs.~\cite{Kobayashi:2012kh,Kobayashi:2014wsa} by allowing the scalar
field to depend on the radial {\em and time} coordinates, since a
nontrivial time dependence of $\phi$ allows for the existence of hairy
BHs \cite{Babichev:2013cya}. Following
Refs.~\cite{Babichev:2013cya,Kobayashi:2014eva}, we assume the scalar
field to have the functional form:
\begin{equation}\label{phipsi}
\phi=\phi(t,r)=q t+\psi(r)\ .
\end{equation}
Then the kinetic energy $X$ is independent of $t$:\
\begin{equation}\label{Xr}
X=X(r)=\frac{1}{2}\left(\frac{q^2}{A(r)}-B(r)\psi'^2\right)\ ,
\end{equation}
where the prime means differentiation with respect to the radial
coordinate $r$.
Then in \eqref{eab},
$X_{\alpha\beta}=-\psi'^2\delta^{r}{_{\alpha}}\delta^{r}{_{\beta}}/2-q^2/2\delta^{t}{_{\alpha}}\delta^{t}{_{\beta}}$.
The $tt$ and $rr$ components of Eq. \eqref{eab}
yield two equations
\begin{align}
{\cal E}_{tt}&=0\ , \label{EqA}\\
{\cal E}_{rr}&=0\, , \label{EqA2}
\end{align}
and
the scalar field equation of motion \eqref{ep}
in the background metric \eqref{Hartle2}
is given by
\begin{equation}
{\cal E}_{\phi}=0\,, \label{Eqphi}
\end{equation}
where the explicit form of the left-hand sides of \eqref{EqA}-\eqref{Eqphi}
is quite lengthy, and it can
be found in Appendix~\ref{app:FE}.
For a static scalar field ($q=0$), Eqs.~(\ref{EqA})-(\ref{Eqphi})
reproduce the results obtained
in~\cite{Kobayashi:2012kh,Kobayashi:2014wsa}; for reflection-symmetric
theories, they reduce to the results of~\cite{Kobayashi:2014eva}.

For slowly rotating solutions at linear order in the BH angular
velocity, the only nonvanishing component of the equations of motion
yields a second-order ordinary differential equation for the variable
$\omega(r)$:
\begin{equation}\label{Eqo}
{\cal E}_{t\varphi}=0 \,.
\end{equation}
Again, the explicit form of the left-hand side can be found in
Appendix~\ref{app:FE}.

Taken together, Eqs.~(\ref{EqA})--(\ref{Eqo}) provide a full
description of vacuum space-times at linear order in rotation. We now
consider two special cases as sanity checks of the equations of
motion.


\subsection{General relativity}

As stated in the introduction, the Einstein-Hilbert Lagrangian of GR
corresponds to setting $G_4=1/2$ and all the other functions equal to
zero.  In this case the equation of motion for the function
$\omega(r)$ simply reads
\begin{equation}
\omega''+\frac{\omega'}{2}\left(\frac{B'}{B}+\frac{8}{r}-\frac{A'}{A}\right)=0\ ,
\end{equation}
in agreement with the frame-dragging equation found by Hartle
\cite{Hartle:1967he}.
If the nonrotating background is the Schwarzschild solution this
further simplifies to
\begin{equation}\label{FDGR}
\omega''+\frac{4}{r}\omega'=0\ .
\end{equation}

\subsection{Einstein-dilaton-Gauss-Bonnet gravity}
EdGB gravity \cite{Kanti:1995vq} corresponds to
the choice of Eqs.~(\ref{EDGBK})-(\ref{EDGBG5}).
If the coupling is linear in the field -- i.e. $\xi(\phi)= \alpha\phi$ as
in~\cite{Sotiriou:2013qea,Sotiriou:2014pfa}, so that the theory is
shift symmetric -- and $q=0$, we get
\begin{align}
(8\alpha B\phi'&-r)\omega''+\left[12\alpha\phi'B'+8\alpha B\phi''+\frac{24\alpha}{r}B\phi'\right.\nonumber\\
&\left.-4\alpha B\phi'\frac{A'}{A}+\frac{r}{2}\left(\frac{A'}{A}-\frac{B'}{B}\right)-4\right]\omega'=0 \ .
\end{align}
If instead we use an exponential coupling of the form $\xi=e^\phi$ and
we set $q=0$, the frame-dragging equation becomes
\begin{align}
&\omega''\left(\frac{2}{B}r^2-2re^\phi
  \phi'\right)+\frac{\omega'r}{B}\left(8-r
  \frac{A'}{A}+\frac{B'}{B}r\right) \nonumber \\
&-\omega'e^\phi\left[2\phi''r+6\phi'+r\phi'\left(3\frac{B'}{B}+2\phi'-\frac{A'}{A}\right)\right]=0\ ,
\end{align}
in agreement with the result of Ref.~\cite{Pani:2009wy}.

\section{Nonminimal derivative coupling to the Einstein tensor}
\label{sec:nonmin}

In this section we apply the formalism derived above to rotating
solutions in a class of Horndeski theories characterized by a
nonminimal derivative coupling with the Einstein tensor of the
form~\eqref{nonminact}.  The theory defined by this action is
invariant under both shift symmetry ($\phi\to \phi+c$) and reflection
symmetry ($\phi\to -\phi$). Shift symmetry allows us to write the
equation of motion for the scalar field $\phi$ as a current
conservation equation \cite{Babichev:2013cya,Sotiriou:2013qea,Sotiriou:2014pfa}:
\begin{equation}\label{concurrent}
\nabla_{\mu}J^{\mu}=0\ .
\end{equation}
In particular, for the action (\ref{nonminact}), the conservation
equation~\eqref{concurrent} reduces to
\begin{equation}\label{phiegG}
(\eta g^{\mu\nu}-\beta G^{\mu\nu})\nabla_{\mu} \partial_{\nu}\phi =0\ .
\end{equation}
Moreover, following \cite{Kobayashi:2014eva} we shall parametrize our
solutions in terms of three auxiliary functions:
\begin{align}\label{aux}
\Lambda&=-\frac{\eta}{\beta}\ ,\\
{\cal F}(X)&=-\frac{-2X\beta \eta+\zeta\eta+\beta\Lambda_0}{2X\beta^2}\ ,\\
{\cal G}(X)&=2(\zeta-\beta X)\ .
\end{align}
Using this parametrization, BH configurations within this theory can
be easily obtained with the following procedure. The $tt$ component of
the equations of motion, Eq.~(\ref{EqA}), leads to the equation
\begin{equation}
-\frac{2A(r)^2}{q^2r{\cal G}}\frac{d}{dr}[X{\cal G}(1-r^2{\cal F}(X))]=0\ ,
\end{equation}
which can be integrated with the solution
\begin{equation}\label{Xalg}
X{\cal G}^2(X)[1-r^2{\cal F}(X)]=C\ ,
\end{equation}
where $C$ is a constant. Equation~(\ref{Xalg}) determines $X(r)$
algebraically. Then the metric function $A(r)$ can be found by solving
Eq.~(\ref{EqA2}), which yields
\begin{equation}\label{Eq2}
(rA)'=\frac{q^2}{2X}\frac{1-r^2\Lambda}{1-r^2{\cal F}(X)}\ .
\end{equation}
Finally, the metric function $B(r)$ can be found from Eq.~(\ref{concurrent}):
\begin{equation}\label{Eq3}
B(r)=\frac{2X}{q^2}[1-r^2 {\cal F}(X)]A(r)\ .
\end{equation}
With the choice (\ref{aux}), the frame-dragging equation for
$\omega(r)$ has a particularly simple form:
\begin{equation}\label{omegagG}
{\cal G}\omega''+\omega'\left[{\cal G}_{X} X’+\frac{1}{2}\left(\frac{8}{r}-\frac{A'}{A}+\frac{B'}{B}\right){\cal G} \right]=0 \,.
\end{equation}

As an extension of Ref.~\cite{Kobayashi:2014eva}, we now consider
nonrotating BH solutions of Eqs.~(\ref{Xalg})-(\ref{Eq3}) in different
subcases and investigate the slow-rotation corrections predicted by
Eq.~(\ref{omegagG}) for each of these solutions.

\paragraph*{Case 1: ${\cal F}=0$.}
One possibility to satisfy Eq.~(\ref{Xalg}) is to impose ${\cal F}(X_f)=0$,
where following the notation of \cite{Kobayashi:2014eva} we define
$X_f$ to be the value of $X$ for which ${\cal F}(X_f)=0$, and
$C=X_f{\cal G}^2(X_f)$. In this case, the metric components and the
scalar field read
\begin{align}
A(r)=&-\frac{\mu}{r}+\frac{q^2}{2X_f}\left(1+\frac{\eta}{3\beta}r^2\right)\ ,\label{case11}\\
B(r)=&\frac{2X_f}{q^2}A(r)\ ,\label{case12}\\
\psi'(r)^2=&\frac{q^2-2X_fA(r)}{A(r)B(r)}\label{case13}\ ,
\end{align}
where $\mu$ is an integration constant.

With a rescaling of the time variable $q^2=2X_f$,
Eqs.~(\ref{case11})-(\ref{case13}) represent a BH solution with an
effective cosmological constant $\Lambda=-\eta/\beta$ and a nontrivial
profile for the scalar field. Replacing this solution into
Eq.~(\ref{omegagG}) we find that $\omega(r)$ satisfies the same equation
\eqref{FDGR} as in GR. The standard solution of this equation is
\begin{equation}
\omega=c_1+\frac{c_2}{r^3}\ ,
\end{equation}
where $c_1$ and $c_2$ are integration constants which can be fixed by
imposing appropriate boundary conditions.

\paragraph*{Case 2: ${\cal G}=0$.}

Another class of solutions of Eq.~(\ref{Xalg}) corresponds to choosing
${\cal G}(X_{\cal G})=0$. In this case, from Eq.~(\ref{omegagG}) we
see that the coefficients of both $\omega''$ and $\omega'$ vanish, and
there are no corrections at linear order.

\paragraph*{Case 3: $q=0$.}
Finally, we consider the case in which the scalar field is
time independent ($q=0$). Integration of the equations of motion for
$A(r)$ and $B(r)$ leads to \cite{Minamitsuji:2013ura}

\begin{align}
A(r)=&\frac{1}{12 \beta  \zeta ^2 \eta ^2 r}\Big\{r (\zeta  \eta -\beta  \Lambda_0) \left[\zeta  \eta  \left(9 \beta +\eta
   r^2\right)\ \right.\nonumber\\
   &+\left.\beta  \Lambda_0 \left(3 \beta -\eta  r^2\right)\right]-24 \beta  \zeta
   ^2 \eta ^2 \mu\Big\}\ \nonumber\\
&+\frac{\sqrt{\beta } (\beta  \Lambda_0
    +\zeta  \eta )^2 \arctan\left(\frac{\sqrt{\eta } r}{\sqrt{\beta }}\right)}{4 \zeta ^2
   \eta ^{5/2} r}\ ,\\
B(r)=&\frac{4 \zeta ^2 \left(\beta +\eta  r^2\right)^2}{\left(2 \beta  \zeta -\beta  \Lambda_0 r^2+\zeta  \eta  r^2\right)^2}A(r)\ ,
\end{align}
where again $\mu$ is an integration constant,
while for the scalar field we obtain:
\begin{equation}
\psi'(r)^2=-\frac{(\beta  \Lambda_0 +\zeta  \eta ) \left[r^3 (\zeta  \eta -\beta  \Lambda_0)+2 \beta  \zeta  r\right]^2}{4 \beta  \zeta ^2 \left(\beta +\eta  r^2\right)^3A(r)}\ .
\end{equation}
Replacing the former expressions into Eq.~\eqref{omegagG}, we find
that the frame-dragging function $\omega(r)$ satisfies once again the
same equation \eqref{FDGR} as in GR.

\section{Why the baldness?}
\label{sec:current}

The no-hair theorems for static, spherically symmetric BHs proved in
Refs.~\cite{Hui:2012qt,Sotiriou:2013qea,Sotiriou:2014pfa} rely
crucially on shift symmetry, which allows us to write the equation of
motion for $\phi$ as the conservation equation~\eqref{concurrent}.
In this section we discuss how these theorems can be generalized to
the case where we consider first-order rotational corrections and
time-dependent scalar fields of the form~\eqref{phipsi}.  In this
case, we can show that the nontrivial components of $J^\mu$ are given
by
\begin{widetext}
\begin{align}
J^{r}=&\phantom{+}B\psi'\bigg[-G_{2\tn{X}}+\frac{B\psi'}{2}\left(\frac{A'}{A}+\frac{4}{r}\right)G_{3\tn{X}}+\frac{2B}{r}\left(\frac{A'}{A}+\frac{B-1}{Br}\right)G_{4\tn{X}}
-\frac{2B^2\psi'^2}{r}\left(\frac{A'}{A}+\frac{1}{r}\right)G_{4\tn{XX}}\nonumber\\
&-\frac{B\psi'}{2r^2}\frac{A'}{A}(3B-1)G_{5\tn{X}}+\frac{A'}{A}\frac{B^3\psi'^3}{2r^2}G_{5\tn{XX}}\bigg]
+\frac{q^2}{A}B\psi'\left[\frac{2B}{r}\frac{A'}{A}G_{4\tn{XX}}-\frac{B^2\psi'}{2r^2}\frac{A'}{A}G_{5\tn{XX}}\right]\nonumber\\
&+\frac{q^2}{A}\left[-\frac{B}{2}\frac{A'}{A}G_{3\tn{X}}+\frac{B}{2r^2}\frac{A'}{A}(B-1)G_{5\tn{X}}\right]\ ,\label{Jr}
\end{align}

\begin{align}
\frac{A}{q}J^{t}=&\phantom{+}G_{2\tn{X}}-\left[B\psi''+\frac{B}{2}\left(\frac{B'}{B}+\frac{A'}{A}+\frac{4}{r}\right)\psi'\right]G_{3\tn{X}}
-\frac{2}{r}\left(B'+\frac{B-1}{r}\right)G_{4\tn{X}}+\frac{2B^2\psi'}{r}\left[2\psi''\right.\nonumber\\
&\left.+\left(\frac{B'}{B}+\frac{A'}{A}+\frac{1}{r}\right)\psi'\right]G_{4\tn{XX}}+\frac{B}{r^2}\left[(B-1)\psi''+\frac{1}{2}\left(\frac{A'}{A}B-\frac{B'}{B}-\frac{A'}{A}+3B'\right)\psi'\right]G_{5\tn{X}}\nonumber\\
&-\frac{B^3\psi'^2}{2r^2}\left[2\psi''+\left(\frac{A'}{A}+\frac{B'}{B}\right)\psi'\right]G_{5\tn{XX}}\nonumber\\
=&-\frac{J^r}{B\psi'}-\frac{2A}{r}\left[\left(G_{4\tn{X}}-\frac{B\psi'}{2r}G_{5\tn{X}}\right)\left(\frac{B}{A}\right)'
+\left(2G_{4\tn{X}}'+\frac{B\psi'}{2r}G_{5\tn{X}}'-\frac{G_{5}'}{2r\psi'}\right)\frac{B}{A}-\frac{1}{2A\psi'}\left(rG_{3}'+\frac{G_{5}'}{r}\right)\right]\ .\label{Jt}
\end{align}
\end{widetext}
For shift-symmetric theories, $G_i=G_i(X)$.  These expressions can be
used to extend the no-hair theorems of
Refs.~\cite{Hui:2012qt,Sotiriou:2013qea,Sotiriou:2014pfa} to the cases
considered in this paper.

For clarity and completeness, let us begin with a short summary of the
original proof given in \cite{Hui:2012qt} (with the amendments of
Refs.~\cite{Sotiriou:2013qea,Sotiriou:2014pfa}).

\subsection{A review of the no-hair theorem for nonrotating black
  holes with a time-independent scalar field}

The no-hair theorem of Ref.~\cite{Hui:2012qt} applies to static,
spherically symmetric, asymptotically flat solutions in
shift-symmetric theories. It consists of the following line of
reasoning:
\begin{enumerate}
\item Assuming that the scalar field $\psi(r)$ has the {\em same
    symmetries as the metric} (the time-dependent scalar field
  of~\cite{Babichev:2013cya} obviously violates this first assumption),
  the only nonvanishing component of $J^{\mu}$ for a spherically
  symmetric background is $J^r$, i.e. $J^\mu=(J^r,0,0,0)$.
\item Given a spherically symmetric space-time, defined by the line
  element \eqref{Hartle2} with $\omega(r)=0$,
we require $J^2=J^{\mu}J_{\mu}$ to remain finite at the horizon $r_h$.
Since
\begin{equation}
  J^2=\frac{(J^r)^2}{B}
\end{equation}
and $B\to 0$ for $r\to r_h$, this regularity condition
implies that $J^r=0$ at the horizon.
\item For a spherically symmetric space-time, the conservation
  equation~\eqref{concurrent} reduces to
\begin{equation}\label{eqJr}
\frac{1}{\sqrt{-g}}\partial_{\mu}(\sqrt{-g}J^\mu)=\partial_r J^r+\frac{2}{r}J^r=0\ ,
\end{equation}
which can be easily integrated. The solution is $J^{r} r^2=K$, where
$K$ is an integration constant. At the horizon the areal radius $r$
cannot be zero. This implies that $K=0$, and therefore that
\begin{equation}\label{Jcond}
J^r=0\quad \forall\ r\ .
\end{equation}
\item The current $J^r$ can be schematically written as
\begin{equation}\label{Jrschematic}
J^r=B \psi' F(g,g',g'',\psi')\ ,
\end{equation}
where $F$ is a generic function of the metric, its first and second
derivatives, and $\psi'$. At spatial infinity, asymptotic flatness
implies that $B\to 1$ and $\psi'\to 0$, while $F$ tends to a nonzero
constant. This last condition is dictated by the requirement that the
scalar field's kinetic energy should have the standard form: in the
weak-field limit, the action contains a term that is quadratic in the
field derivatives and $J_\mu\to \partial_{\mu}
\phi$, up to an overall
constant of normalization.  If we now move ``inward'' towards the
horizon, by continuity $F$ and $B$ will still be nonzero, and
therefore $J^r\neq 0$, which contradicts Eq.~(\ref{Jcond}). This
contradiction can be avoided if $\psi'=0$ for any choice of $r$, which
fixes $\psi={\rm constant}$ or (without loss of generality, since the
theory is shift symmetric) $\psi=0$.
\end{enumerate}

Sotiriou and Zhou~\cite{Sotiriou:2013qea,Sotiriou:2014pfa} pointed out
a loophole in the last step of this proof.  For Horndeski gravity
theories with shift symmetry, the conserved current can be written as
\begin{align}
J^r=&-BG_{2\tn{X}}\psi'+\frac{B^2\psi'^2}{2}\left(\frac{A'}{A}+\frac{4}{r}\right)G_{3\tn{X}}  \nonumber\\
&+\frac{2B^2\psi'}{r}\left(\frac{A'}{A}-\frac{1}{Br}+\frac{1}{r}\right)G_{4\tn{X}}\ \nonumber\\
&-\frac{2B^3\psi'^3}{r}\left(\frac{A'}{A}+\frac{1}{r}\right)G_{4\tn{XX}}  \nonumber\\
&-\frac{B^3\psi'^2}{2r^2}\frac{A'}{A}\left(\frac{3B-1}{B}\right)G_{5\tn{X}}+\frac{A'}{A}\frac{B^4\psi'^4}{2r^2}G_{5\tn{XX}}\ .\label{Jrbis}
\end{align}
Depending on the particular form of the coupling functions $G_i$ we
have essentially two options:\footnote{A third case where $J^r$
  contains negative powers of $\psi'$ can be excluded because it
  generally corresponds to theories that would not admit flat space
  with a trivial scalar configuration as a solution, leading to
  violations of local Lorentz symmetry~\cite{Sotiriou:2014pfa}.}
\begin{itemize}
\item[(a)] $J^r$ depends linearly on $\psi'$. This is the case considered
  in Ref.~\cite{Hui:2012qt}, for which $F\to -G_{2X}$ as
  $r\to \infty$.
\item[(b)] $J^r$ contains terms which are independent of $\psi'$, but no
  negative powers of $\psi'$.
\end{itemize}
This second case represents a loophole for the no-hair theorem of
Ref.~\cite{Hui:2012qt}. Indeed, in this case the asymptotic behavior
of $F$ is not trivially determined.

This is illustrated most clearly by looking at two specific examples:
EdGB gravity and theories with nonminimal derivative coupling to the
Einstein tensor.

In the first case the conserved current reduces to
\begin{equation}\label{JrEDGB}
J^r_{\tn{EdGB}}=-B \psi'- 4\alpha \frac{A'B(B-1)}{Ar^2}\ ,
\end{equation}
where we specialized to a linear coupling function $\xi(\phi)=\alpha\phi$ in
Eq.~\eqref{EdGBaction}, so that the theory becomes shift symmetric
(recall that the Gauss-Bonnet combination is a topological invariant).
The current \eqref{JrEDGB} contains a term independent of $\psi'$ as
in case (b) above, corresponding to the loophole pointed out in
Refs.~\cite{Sotiriou:2013qea,Sotiriou:2014pfa}. The current vanishes
at infinity, but for smaller radii the choice of $F$ is nontrivial and
leads to scalar hair growth.

For the nonminimal derivative coupling theory we have instead
\begin{equation}\label{JrGg}
J^r_{\tn{Gg}}=B\psi'\left[-2\eta+\frac{2B}{r}\left(\frac{A'}{A}-\frac{1}{Br}+\frac{1}{r}\right)\beta\right]\ .
\end{equation}
This expression for the current falls into case (a) above. The current
depends linearly on $\psi'$, $F\to -2\eta$ for $r\to\infty$, and $F$
stays finite even at finite radii by continuity, as required by the
arguments of~\cite{Hui:2012qt}, so we are forced to set $\psi'=0$ and
$\psi$ is a constant, which can be set to zero.
Asymptotic flatness was of course a key ingredient in these
arguments. Hairy solutions in theories with nonminimal derivative
coupling are {\it not} asymptotically flat (see
e.g.~\cite{Rinaldi:2012vy,Minamitsuji:2013ura,Anabalon:2013oea}).

\subsection{Extension to slow-rotation and time-dependent scalar fields}

What is crucial for the present work is that the arguments above apply
also to rotating BH solutions at linear order in rotation. This is
because, as argued in Ref.~\cite{Sotiriou:2013qea}, the scalar field
$\phi$ (like all scalar quantities) is affected by rotation only at
second order, and therefore the expression (\ref{Jr}) for the current
$J^r$ remains unchanged at linear order. Similarly, $J^{\theta}$ is
still equal to zero at linear order. The component $J^{\varphi}$
acquires a nonzero value proportional to the BH angular momentum;
however $J^{\varphi}$ is independent of $\varphi$, and therefore it
does not contribute to the current conservation
equation~(\ref{concurrent}).

At first sight, the fact that no-hair theorems still hold true at
linear order in rotation even for time-dependent scalar fields may be
surprising. However this no-hair property can be proved through a
simple extension of the arguments valid for static, nonrotating
solutions. Let us extend the original argument to theories
with
time-dependent scalar fields of the form~\eqref{phipsi}:

\begin{enumerate}
\item When $\phi$ has the form~\eqref{phipsi} the current has a
  nonzero time component, i.e. $J^{\mu}=(J^r,0,0,J^t)$, and its norm
  becomes
\begin{equation}
J^2=\frac{(J^r)^2}{B}-(J^t)^2 A\ .
\end{equation}

\item By imposing regularity at the horizon, where $A\to 0$, $B\to 0$,
  we conclude that $J^r\to 0$ as $r\to r_h$. This is true as long as
  $J^t$ does not diverge in the limit $r\to r_h$, i.e., as long as the
  quantity in square brackets in the last line of Eq.~\eqref{Jt} is
  finite. For reflection-symmetric theories ($G_3=G_5=0$), this latter
  requirement simplifies to the condition that $(B/A)'$ should be
  finite~\cite{Kobayashi:2014eva}.

\item
In principle, the current conservation equation~(\ref{eqJr})
acquires an extra term because $J^t\neq 0$:
\begin{equation}
\partial_r J^r+\frac{2}{r}J^r+\partial_tJ^t=0\ .
\end{equation}
However Eq.~\eqref{Jt} shows that in the present case $J^t$ is
independent of time, so this term vanishes: $\partial_t J^t=0$.
Following the reasoning below Eq.~\eqref{eqJr}, we conclude that
$J^r=0$ for all $r$ even for scalar fields with a linear time
dependence.  Note that for a time-dependent scalar field, in general,
the $tr$ component of the gravitational equations ${\cal E}_{tr}=0$
may be nontrivial, indicating the existence of an energy flux in the
radial direction.  However Ref.~\cite{Babichev:2015rva} showed that,
for the linear-in-time ansatz \eqref{phipsi}, ${\cal E}_{tr}$ is
proportional to $J^r$ under the assumptions of diffeomorphism
invariance and shift symmetry.  Therefore the condition $J^r=0$ always
ensures that ${\cal E}_{tr}=0$: the linear time dependence
\eqref{phipsi} does not give rise to an energy flux in the radial
direction.

\item The current \eqref{Jr} has the form \eqref{Jrschematic}, where
  $F(g,g',g'',\psi')$ is an unspecified function. This allows us to
  borrow in its entirety the reasoning of
  Ref.~\cite{Sotiriou:2014pfa}. We can exclude cases where $J^r$
  contains negative powers of $\psi'$. When all terms in $J^r$ contain
  positive powers of $\psi'$, $\psi'=0$ for all $r$ and the no-hair
  theorem of~\cite{Hui:2012qt} applies. The only exception is the case
  where $J^r$ contains one or more terms with no dependence on
  $\psi'$, but no terms with negative powers of $\psi'$; and then,
  following Sec.~IIB of~\cite{Sotiriou:2014pfa}, shift symmetry and
  Lovelock's theorem imply that the action must contain a term
  proportional to the Gauss-Bonnet invariant.
\end{enumerate}

This generalized no-hair theorem can be used to justify the absence of
corrections to GR at linear order that we found in
Sec.~\ref{sec:nonmin}.
For a theory with nonminimal derivative coupling to the Einstein
tensor, the nonzero components of the current can be obtained by
specializing Eqs.~\eqref{Jr}-\eqref{Jt}, with the result
\begin{align}
  J^r_{\tn{Gg}}&=B\psi'\left[-2\eta+
                  \frac{2B}{r}\left(\frac{A'}{A}-\frac{1}{Br}+\frac{1}{r}\right)\beta\right]\ ,\label{JrqGg}\\
  \frac{A}{q}J^t_{\tn{Gg}}&=2\eta+\frac{2B}{r}\left(\frac{1-B}{Br}-\frac{B'}{B}\right)\beta\ .\label{JtqGg}
\end{align}
The $J^r$ component is identical to the static case of
Eq.~\eqref{JrGg}, it does not contain any $\psi'$-independent terms,
and the no-hair theorem of~\cite{Hui:2012qt} implies that
asymptotically flat solutions must be the same as GR.

In conclusion, the only no-hair violations at linear order in rotation
when the scalar field depends linearly on time and when we require
asymptotic flatness can occur in one of two cases:

\begin{itemize}
\item[(i)] if the scalar field has a linear coupling to the
  Gauss-Bonnet invariant, or

\item[(ii)] if, as proposed in Ref.~\cite{Babichev:2013cya}, the field
  equations of the theory guarantee that the current vanishes
  identically ($J^r=0$) because $F(g,g',g'',\psi')=0$ as a consequence
  of the field equations. Note that this is only possible for special
  forms of the functions $G_i$, and that the scalar field must then be
  time dependent (i.e., it must violate some of the symmetries of the
  metric) in order to be regular at the horizon.
\end{itemize}

\section{Conclusions}
\label{sec:concl}

In this work we studied leading-order rotational corrections to a
broad class of BH solutions in Horndeski gravity. With the known
exception of EdGB
gravity~\cite{Kanti:1995vq,Pani:2009wy,Sotiriou:2013qea,Sotiriou:2014pfa,Ayzenberg:2014aka,Maselli:2015tta},
we have found that the frame-dragging function $\omega(r)$, which
describes the leading-order rotational corrections, is exactly the
same as in GR for all of the Horndeski BH solutions known in the
literature. This result applies even to asymptotically flat solutions
that violate the no-hair theorems by requiring the scalar field to be
time dependent (so that the scalar field does not respect the same
symmetries as the metric), as proposed in
Ref.~\cite{Babichev:2013cya}.

The formalism developed in this paper can be extended in various
directions. First of all, the no-hair theorem proved in
Sec.~\ref{sec:current} at first order in rotation is not expected
to hold at second order, where the continuity equation will be
modified.
Calculations of BH solutions at second order in rotation,
along the lines of~\cite{Ayzenberg:2014aka,Maselli:2015tta}, are
already underway~\cite{MaselliPrep}.

Even for nonrotating Horndeski BHs, studies of stability and
perturbative dynamics (as encoded in their quasinormal mode spectrum:
see e.g.~\cite{Berti:2009kk} for a review) are still in their
infancy. One of us~\cite{Minamitsuji:2014hha} studied massless scalar
field perturbations of static BH solutions in theories with field
derivative coupling to the Einstein tensor. More in general,
gravitational perturbations of static, nonrotating space-times can be
explored using the formalism developed in
Refs.~\cite{Kobayashi:2012kh,Kobayashi:2014wsa}. The present work lays
the foundations to study quasinormal modes and look for super-radiant
instabilities using the slow-rotation perturbative techniques
reviewed, e.g., in Ref.~\cite{Pani:2013pma}.

Another important extension concerns compact stars in Horndeski
gravity. Slowly rotating compact stars in EdGB gravity were studied
in~\cite{Pani:2011xm}. Cisterna {\it et al.}~\cite{Cisterna:2015yla}
investigated compact objects in theories with a nonminimal derivative
coupling of the scalar field with the Einstein tensor. Our formalism
can be extended relatively easily to study compact stars in broader
classes of Horndeski gravity, and to understand whether genuine
strong-field deviations from GR (similar to the ``spontaneous
scalarization'' phenomena proposed by Damour and
Esposito-Far\`ese~\cite{Damour:1993hw}) can occur in some sectors of
the Horndeski gravity action, see e.g.~\cite{Chen:2015zmx} for recent
work in this direction.


\acknowledgments

We thank Eugeny Babichev, Vitor Cardoso, Sante Carloni, Adolfo
Cisterna, Leonardo Gualtieri, Matteo Lulli, Paolo Pani and Eleftherios
Papantonopoulos for discussions.
A.M.  was supported by NSF Grants No.~1205864, No.~1212433 and No.~1333360.
E.B. was supported by NSF CAREER Grant No.~PHY-1055103 and by FCT
Grant No.~IF/00797/2014/CP1214/CT0012 under the IF2014 program.
H.O.S was supported by NSF CAREER Grant No.~PHY-1055103 and by a
summer research assistantship award from the University of
Mississippi. E.B. and H.O.S. thank the Instituto Superior T\'ecnico
(Lisbon, Portugal), where part of this project was completed, for the
hospitality.
M.M. was supported by the FCT Portugal through Grant
No.~SFRH/BPD/88299/2012.

\appendix

\section{Field equations}
\label{app:FE}

In this appendix we list the left-hand side of the field
equations. For clarity, we split all of the left-hand sides of the
field equations as a sum of two contributions, so that the case of
time-independent scalar fields can more easily be recovered by setting
$q=0$:
\begin{align}
{\cal E}_{\alpha\beta}={\cal E}_{\alpha\beta}^{(0)}+\frac{q^2}{A}{\cal
  E}_{\alpha\beta}^{(t)}\ ,\\
{\cal E}_{\phi}={\cal E}_{\phi}^{(0)}+\frac{q^2}{A}{\cal
  E}_{\phi}^{(t)}\ .
\end{align}
Let us remark that the equations of motion still depend on the
specific form of the $G_i$'s, which are functions of the kinetic
energy \eqref{Xr}, and therefore may contain $q$-dependent terms;
therefore we must evaluate all of the functions $G_{i}$ at $q=0$ to recover
the time-independent limit.  The explicit forms of the various terms
are
\begin{widetext}
\begin{align}
{\cal E}_{tt}^{(0)}=&\phantom{+}G_2+B  \psi'^2G_{3\phi} -\frac{B\psi'^2}{2}\left(  B' \psi' +2\,B \psi''  \right) G_{{3\tn{X}}}-\frac{2}{r}\left( \frac{B -1}{r}+B'  \right)G_{4}
 -\frac {2B^2\psi'}{r}  \left( \frac{\psi'}{r} +2\frac{B'}{B} \psi' +2\psi''  \right) G_{4\tn{X}}\nonumber\\
&-B\left( \frac{4}{r}\psi' +\frac{B'}{B} \psi' +2 \psi''  \right) G_{4\phi}+\frac{2 B^2 \psi'^3}{r}\left(  B' \psi' +2\,B \psi''  \right) G_{4\tn{XX}}
-B^2\psi'^2\left( \frac{4}{r} \psi' -\frac{B'}{B} \psi' -2 \psi''  \right) G_{4\tn{X}\phi}\nonumber\\
&-2B  \psi'^2G_{4\phi \phi}+\frac{B  \psi'^2}{2r^2} \left( 5B' B \psi' +6B^2\psi'' - B' \psi' -2B \psi''  \right) G_{5\tn{X}}+\frac { B^3 \psi'^3}{r} \left(\frac{\psi'}{r}-\frac{B'}{B} \psi' -2\psi''\right) G_{{5\tn{X}\phi }}\nonumber\\
& -\frac{B^{3} \psi'^{4}}{2r^2}\left(  B' \psi' +2\,B \psi''  \right) G_{5\tn{XX}}+\frac {B  \psi'}{r}  \left( 3B'\psi' +4 B\psi'' +\frac{\psi'}{r} +B \frac{\psi'}{r}  \right) G_{5\phi}
+\frac {2B^2 \psi'^3}{r}G_{5\phi \phi}\ ,
\end{align}
\begin{align}
{\cal E}_{tt}^{(t)}=&-G_{2\tn{X}}+G_{3 \phi}+\frac{B}{2} \left( 4\frac{ \psi'}{r} +\frac{ B'}{B}\psi' +2 \psi''  \right) G_{3\tn{X}}+\frac{2}{r} \left( \frac{B -1}{r}+B'  \right) G_{4 \tn{X}}
-\left( \frac{4 B}{r} \psi' + B' \psi' +2 B \psi''  \right) G_{4 \tn{X}\phi}\nonumber\\
&-\frac {2B  \psi'}{r}  \left( B' \psi' +2 B \psi'' +\frac{B}{r} \psi'  \right) G_{4 \tn{XX}}
 -\frac{1}{2r^2} \left(3  B' B \psi' - B' \psi' -2 B \psi'' +2  B^2\psi''  \right) G_{5 \tn{X}}
- \frac{1}{r}\left(\frac{ B -1}{r}+B'  \right) G_{5 \phi}\nonumber\\
 &   +\frac {B  \psi'}{r}  \left( B' \psi' +2 B \psi'' +\frac{B}{r} \psi'  \right) G_{5 \tn{X}\phi} +\frac { B^2 \psi'^{2}}{2r^2} \left(  B' \psi' +2 B \psi''  \right) G_{5\tn{ XX}}\ ,
\end{align}

\begin{align}
{\cal E}_{rr}^{(0)}=&\phantom{+}G_{2}+B\psi'^2G_{2\tn{X}}-B  \psi'^2G_{3\phi}
-\frac { B^2 \psi'^3}{2} \left( \frac{4}{r}+ \frac{A'}{A}\right) G_{3\tn{X}}-\frac{2}{r} \left( B\frac{A'}{A}+\frac{B-1}{r}\right) G_{4}
-B  \psi'  \left( \frac{4}{r} + \frac{A'}{A}\right) G_{4\phi}\nonumber\\
& -\frac {2B  \psi'^2}{r} \left( 2B \frac{A'}{A} +\frac{2B-1}{r}  \right) G_{4\tn{X}}+B^2 \psi'^{3} \left( \frac{4}{r}+ \frac{A'}{A}\right) G_{4\tn{X}\phi}+\frac{2B^{3}\psi'^{4}}{r}\left(\frac{A'}{A}+\frac{1}{r}\right)G_{4\tn{XX}}\nonumber\\
&+\frac{B\psi'^2}{r} \left(3B\frac{A'}{A}+\frac{3B-1}{r}\right) G_{5\phi}+\frac { B^2\psi'^3 }{2r^2} \frac{A'}{A}\left( 5\,B -1 \right) G_{5\tn{X}}-\frac{B^3\psi'^{4}}{r} \left(\frac{A'}{A}+\frac{1}{r}\right) G_{5\tn{X}\phi}-\frac{B^4\psi'^{5}}{2r^2}\frac{A'}{A}G_{5\tn{XX}}\ ,
\end{align}
\begin{align}
{\cal E}_{rr}^{(t)}=&-G_{3\phi}+\frac {B  \psi'}{2}\frac{A'}{A} G_{3\tn{X}}+\frac{2B}{r}\frac{A'}{A} G_{4\tn{X}}+2G_{4\phi\phi}-\frac{2B^2\psi'^{2}}{r}\frac{A'}{A} G_{4\tn{XX}}
+B\psi'  \left(\frac{4}{r}-\frac{A'}{A}\right)G_{4\tn{X}\phi}-\frac{2B\psi'}{r}G_{5\phi\phi}\nonumber\\
&-\frac {B \psi'}{2r^2}\frac{A'}{A}\left(3B -1\right)  G_{5\tn{X}}+\frac { B^2 \psi'^{2}}{r} \left(\frac{A'}{A}-\frac{1}{r}\right) G_{5\tn{X}\phi}+\frac{B^3\psi'^{3}}{2r^2}G_{5\tn{XX}}
+\frac{1}{r} \left(\frac{B-1}{r}-B \frac{A'}{A}\right) G_{5\phi}\ ,
\end{align}

\begin{align}
{\cal E}_{t\varphi}^{(0)}=&\phantom{+}\omega G_2 +B \psi'^{2}\omega G_{3\phi}
-\frac{B\psi'^{2}}{2}\omega \left(  B' \psi' +2\,B \psi''  \right) G_{3\tn{X}}
+\frac{B}{2}\left[- \left( \frac{2}{r}\frac{B'}{B}+2 \frac{A''}{A} + \frac{B'}{B}\frac{A'}{A}- \frac{A'^2}{A^2}+\frac{2}{r}\frac{A'}{A}\right) \omega\right.
\nonumber\\
&\left.+\left( \frac{B'}{B}+\frac{8}{r}-\frac{A'}{A}  \right) \omega' + 2\omega''\right] G_{4}- \left[ \left( \frac{ A'}{A} B \psi' +B'  \psi' +2B \psi'' +\frac{2}{r}B  \psi'   \right) \omega -  \omega'   B \psi'  \right]G_{4\phi}\nonumber\\
&+ \frac{B^2\psi'}{2}\left[ -  \left( \frac{2}{r} \frac{A'}{A}\psi'+\frac{4}{r}\frac{B'}{B}\psi' +2\frac{A''}{A}\psi' -\frac{A'^2}{A^2}\psi' +2\frac{B'}{B}\frac{A'}{A}\psi'  +2\frac{A'}{A}\psi'' +\frac{4}{r}\psi''  \right) \omega+2\psi'\omega''\right. \nonumber\\
 &\left.+\left( \frac{8\psi'}{r}  +\frac{2B'}{B} \psi' +2\psi'' - \frac{A'}{A} \psi'  \right) \omega' \right] G_{4\tn{X}}+B^2\psi'^2 \left[\left(\frac{B'}{B} \psi' -\frac{A'}{A}\psi'+2\psi'' -\frac{2}{r}  \psi'   \right) \omega+ \psi'\omega'' \right] G_{4\tn{X}\phi}\nonumber\\
&-2B  \psi'^2\omega G_{4\phi \phi}+\frac{B^2\psi'^3}{2} \left[\left( \frac{2}{r}  \psi' B' + \psi'  B' \frac{A'}{A} +\frac{4}{r}B   \psi' \psi'' +2 B\frac{A'}{A} \psi''  \right) \omega -\left(\psi'B' +2B  \psi''  \right) \omega' \right] G_{4\tn{XX}}\nonumber\\
 &+\frac{B^3\psi'^2}{4r}\left[-\left(5\frac{B'}{B} \psi'- \frac{A'}{A}\psi'  +6 \psi'' +\frac{6}{r} \psi'  \right) \omega'+\left(5\frac{ B'}{B}\frac{A'}{A}\psi'- \frac{A'^2}{A^2} \psi'  +6\frac{A'}{A} \psi'' +2\frac{A''}{A}\psi'  \right) \omega-2\psi'\omega'' \right] G_{5\tn{X}}\nonumber\\
 &+\frac{B^2\psi'^3}{4} \left[ \left( \psi'B' +2B\psi'' -\frac{2B}{r}\psi'\right) \omega' -
\left( \psi' B'\frac{A'}{A} +2B\frac{A'}{A} \psi'' -\frac{2}{r}B \psi' \frac{A'}{A} +\frac{2}{r}\psi'B' +\frac{4}{r}B\psi''  \right) \omega  \right] G_{5\tn{X}\phi}\nonumber\\
& +\frac{B^3 \psi'^4}{4r} \left[    \left(   \psi' B' +2\, B \psi''   \right)\omega'-\frac{A'}{A}   \left(  \psi' B' +2B \psi''  \right) \omega  \right] G_{5\tn{XX}}+ \frac{B\psi'}{4}\left[\left(\frac{A'}{A} B \psi'- \frac{8}{r}B  \psi'  -4B \psi''  -3 B' \psi'  \right) \omega' \right.\nonumber\\
&\left.+ \left( 3B'\frac{A'}{A}\psi' - \frac{BA'^2}{A^2}\psi' +\frac{2}{r}\frac{A'}{A} B  \psi'  +2B\frac{ A''}{ A} \psi' +\frac{6}{r} B' \psi'
+\frac{8}{r}B \psi'' +4\frac{A'}{A} B \psi''  \right) \omega -2 B \psi'\omega''  \right] G_{5\phi}\nonumber\\
&+\frac{B^2\psi'^3}{2} \left[ \left( \frac{2}{r}+ \frac{A'}{A} \right) \omega -\omega'  \right] G_{5\phi\phi}\ ,
\end{align}

\begin{align}
{\cal E}_{t\varphi}^{(t)}=&-\frac{ B  \psi'}{2}\frac{A'}{A}\omega G_{3\tn{X}} -\omega G_{3\phi}  + \left[\left( \frac{2B}{r}\frac{A'}{A}  +2B \frac{ A''}{A} -2B\frac{ A'^{2}}{A^2}+B' \frac{ A'}{A}  \right) \frac{\omega}{2} -\left(  B' +\frac{8B}{r} -2B \frac{A'}{A}  \right)\frac{ \omega'}{2}  -B \omega''  \right] G_{4\tn{X}}\nonumber\\
&+ \left[ \left( 3\frac{A'}{A} B \psi' +\frac{2B}{r}\psi' +B' \psi' +2 B \psi''  \right) \omega- \omega' B \psi'  \right] G_{{4\,X\phi }}+\frac{B}{2} \left[B\psi'  \left(  \frac{B'}{B}  \psi'  +2\psi'' -  \psi' \frac{A'}{A}  \right)\omega'\right.\nonumber\\
&\left.-\frac {A'}{A}B  \psi'  \left(\frac{B'}{B} \psi' -\frac{A'}{A} \psi' +2 \psi'' -\frac{2}{r} \psi'  \right) \omega  \right] G_{4\tn{XX}}+2\omega G_{4\phi \phi}+ \left[ \frac{B^2}{4r}\left( \frac{6\psi'}{r} -3\frac{ A'}{A}  \psi' +3\frac{B'}{B} \psi' +2 \psi''  \right) \omega'\right. \nonumber\\
&\left.-\frac{B^2}{4r}  \left( 3\frac{B'}{B}  \frac{A'}{A}  \psi' +2\frac{A'}{A}  \psi'' +2 \frac{A''}{A} \psi' -3\frac{A'^{2}}{A^2} \psi'  \right) \omega+  \frac {B^2\psi'}{2r} \omega'' \right]G_{5\tn{X}}+ \left[\frac{B^2 \psi' }{4}\left(\frac{A'}{A}\psi' -\frac{ B'}{B} \psi' -2 \psi'' +\frac{2}{r} \psi'   \right) \omega'\right.\nonumber\\
&\left.-\frac{B^2 \psi' }{4} \left(\frac{ A'^{2}}{A^2} \psi' - \frac{B'}{B}\frac{A'}{A}\psi' -2\frac{ A'}{A} \psi'' +\frac{6}{r}\frac{A'}{A}\psi'  +\frac{2}{r}\frac{ B'}{B} \psi' +\frac{4}{r} \psi''  \right) \omega \right] G_{5\tn{X}\phi }\nonumber\\
&+ \frac{B^3\psi'^{2}}{4r}\frac{A'}{A}\left[ \left( \psi' - \frac{AB'}{A'B}\psi' - \frac{2A}{A'}\psi'' \right)\omega' + \left( \frac{B'\psi'}{B} - \frac{A'\psi'}{A} + 2\psi'' \right)\omega \right] G_{5\tn{XX}}\nonumber\\
&+ \frac{B}{4}\left[\left( -\frac{2}{r}\frac{A'}{A}+3 \frac{A'^{2}}{A^2}-2\frac{A''}{A} +\frac{2}{r} \frac{B'}{B} - \frac{B'}{B}\frac{A'}{A}  \right) \omega + \left( \frac{8}{r} + \frac{B'}{B}-3\frac{A'}{A}  \right) \omega' +2\omega'' \right] G_{5\phi}
\nonumber \\
&+ \frac{B}{2}\left[ -\left( 2\frac{ A'}{A} \psi' +\frac{ B'}{B} \psi' +2\psi''+\frac{2}{r}  \psi'  \right) \omega+ \psi'\omega'  \right] G_{5\phi\phi}\nonumber\\
&+\frac{q^2}{A}\frac{B}{4}\frac { A'}{A} \left(\frac {A'}{A}\omega -\omega' \right)\left( G_{5\tn{X}\phi}-
2 G_{4\tn{XX}}+\frac{B\psi'}{r}G_{5\tn{XX}}\right)\ ,
 \end{align}

\begin{align}
{\cal E}_{\phi}^{(0)} &= G_{2\phi} + B \psi'\left( \frac{A'}{2A} + \frac{B'}{2 B} + \frac{2}{r}+\frac{\psi''}{\psi'} \right) G_{2\tn{X}} + B \psi^{\prime 2}\,G_{2\tn{X}\phi} -
B \psi^{\prime 3}\left( \frac{B'}{2} + \frac{B \psi''}{\psi'} \right)G_{2\tn{XX}}
\nonumber \\
&- B \psi'\left( \frac{B'}{B} + \frac{A'}{A} + \frac{4}{r} + \frac{2\psi''}{\psi'} \right) G_{3\phi} -B^2 \psi^{\prime 2} \left( \frac{3 A' B'}{4A B} + \frac{3 B'}{B r} + \frac{A'\psi''}{A\psi'} + \frac{4 \psi''}{\psi' r} + \frac{A''}{2A}-\frac{A^{\prime 2}}{4 A^2} + \frac{2 A'}{Ar} + \frac{2}{r^2} \right) G_{3\tn{X}} \nonumber \\
&+ B^{2} \psi^{\prime 3}\left( \frac{\psi^{\prime\prime}}{\psi^{\prime}}-
\frac{A'}{2 A} + \frac{B'}{2 B}- \frac{2}{r}
\right) G_{3\tn{X}\phi}
+ B^3 \psi^{\prime 4}\left( \frac{B'}{B r} + \frac{A' B'}{4 A B}
+ \frac{2 \psi^{\prime \prime}}{r \psi^{\prime}}
+ \frac{A' \psi^{\prime \prime}}{2 A \psi^{\prime}}
\right) G_{3\tn{XX}} - B \psi^{\prime 2} G_{3\phi\phi}\nonumber \\
&+ B\left(
\frac{A^{\prime 2}}{2 A^2} - \frac{A''}{A} - \frac{A' B'}{2 A B} -
\frac{2 A'}{ A r} - \frac{2 B'}{B r} - \frac{2}{r^2} + \frac{2}{B r^2}
\right) G_{4\phi} \nonumber \\
&+ B^2 \psi' \left(
\frac{A^{\prime 2}}{A^2 r} - \frac{2 A''}{Ar} - \frac{3 A' B'}{A B r}
-\frac{3 A'}{A r^2} - \frac{3 B'}{B r^2} + \frac{A'}{A B r^2}
-\frac{2 A' \psi''}{A \psi' r} - \frac{2 \psi''}{\psi' r^2} + \frac{B'}{B^2 r^2}
+ \frac{2 \psi''}{B \psi' r^2}
\right)G_{4\tn{X}} \nonumber \\
&+B^3 \psi^{\prime 3}
\left(
\frac{2 A''}{Ar} - \frac{A^{\prime 2}}{A^2 r} + \frac{6 A' B'}{A B r}
+\frac{3 A'}{A r^2} + \frac{6 B'}{Br^2} + \frac{8 A' \psi''}{A \psi' r}
+ \frac{8 \psi''}{\psi' r^2} - \frac{B'}{B^2 r^2}
- \frac{2 \psi''}{B \psi' r^2}
\right)G_{4\tn{XX}} \nonumber \\
&+ B^2 \psi^{\prime 2} \left(
\frac{A''}{A} - \frac{A^{\prime 2}}{2 A^2} + \frac{2 A' B'}{A B}
+ \frac{4 A'}{A r} + \frac{8 B'}{B r} + \frac{4}{r^2} + \frac{2}{B r^2}
+ \frac{3 A' \psi''}{A \psi'} + \frac{12 \psi''}{\psi' r}
\right) G_{4\tn{X}\phi}
+ B^2 \psi^{\prime 3}\left(\frac{A'}{A} + \frac{4}{r}\right) G_{4\tn{X}\phi\phi}
\nonumber \\
&+ B^3 \psi^{\prime 4} \left(
\frac{2A'}{Ar} -\frac{2 B'}{Br}- \frac{A' B'}{2AB}  -\frac{A' \psi''}{A \psi'}
-\frac{4\psi''}{\psi' r} + \frac{2}{r^2}
\right) G_{4\tn{XX}\phi}
- B^3 \psi^{\prime 5}\left(
\frac{A' B'}{Ar} + \frac{B'}{r^2} + \frac{2 A' B \psi''}{A \psi' r}
+ \frac{2B \psi''}{\psi' r^2}
\right)G_{4\tn{XXX}} \nonumber\\
&+ B^2 \psi' \left(
\frac{2A''}{Ar} - \frac{A^{\prime 2}}{A^2 r} + \frac{3 A'}{A r^2} +
\frac{3 A' B'}{A B r} + \frac{3 B'}{B r^2} - \frac{A'}{A B r^2} +
\frac{2 A' \psi''}{A \psi' r} + \frac{2 \psi''}{\psi' r^2}
- \frac{B'}{B^2 r^2} - \frac{2 \psi''}{\psi' B r^2}
\right) G_{5\phi} \nonumber \\
&+ B^3 \psi^{\prime 2}\left(
\frac{3 A''}{2 A r^2} - \frac{3 A^{\prime 2}}{4 A^2 r^2}
+ \frac{15 A' B'}{4 A B r^2} + \frac{A^{\prime 2}}{4 A^2 B r^2}
-\frac{A''}{2 A B r^2} - \frac{3 A' B'}{4 A B^2 r^2} +
\frac{3 A' \psi''}{A\psi' r^2} - \frac{A' \psi''}{A B \psi' r^2}
\right)G_{5\tn{X}} \nonumber \\
&+ B^4 \psi^{\prime 4}\left(
\frac{A^{\prime 2}}{4 A^2 r^2} - \frac{A''}{2 A r^2} -\frac{5 A' B'}{2 A B r^2}
- \frac{7 A' \psi''}{2 \psi' A r^2} + \frac{A' B'}{4 A B^2 r^2}
+ \frac{A' \psi''}{2 A B \psi' r^2}
\right) G_{5\tn{XX}} \nonumber \\
&+ B^3 \psi^{\prime 3}\left(
\frac{A^{\prime 2}}{2 A^2 r} - \frac{A''}{A r} - \frac{7 A' B'}{2 A B r}
-\frac{A'}{2 A r^2} - \frac{7 B'}{2 B r^2} -\frac{A'}{2 A B r^2}
-\frac{5 A' \psi''}{A \psi' r} -\frac{5 \psi''}{\psi' r^2}
+\frac{B'}{2 B^2 r^2} + \frac{\psi''}{\psi' B r^2}
\right) G_{5\tn{X}\phi} \nonumber \\
&+ B^2 \psi^{\prime 2} \left(
\frac{A'}{A r} - \frac{1}{B r^2} + \frac{1}{r^2}
\right) G_{5\phi\phi}
- B^3 \phi^{\prime 4} \left(
\frac{A'}{A r} + \frac{1}{r^2}
\right) G_{5\tn{X}\phi\phi} \nonumber \\
&+ B^4 \psi^{\prime 5}\left(
\frac{A' B'}{2 A Br} - \frac{A'}{2 A r^2}
+ \frac{B'}{2Br^2}+
\frac{A' \psi''}{A \psi' r}+ \frac{\psi''}{\psi' r^2}
\right) G_{5\tn{XX}\phi} + B^4 \psi^{6 \prime}\left(
\frac{A' B'}{4 A r^2} + \frac{A' B \psi''}{2 A r^2 \psi'}
\right) G_{5\tn{XXX}}\ ,
\end{align}

\begin{align}
{\cal E}_{\phi}^{(t)} &=  - G_{2\tn{X}\phi}
-
\frac{A' B \psi'}{2A}
G_{2\tn{XX}} + \left( \frac{BA'}{Ar} - \frac{3 B A^{\prime 2}}{4 A^2}
+ \frac{A'B'}{4A} + \frac{B A''}{2A} \right) G_{3\tn{X}} \nonumber \\
&+ G_{3\phi\phi} + B \psi' \left( \frac{B'}{2 B} + \frac{2}{r} + \frac{3 A'}{2 A} + \frac{\psi''}{\psi'} \right) G_{3\tn{X}\phi}
+ B^2 \psi^{\prime 2} \left(
\frac{A'}{Ar} + \frac{A^{\prime 2}}{4 A^2} -\frac{A' \psi''}{2 A \psi'}
-\frac{A' B'}{4 AB} \right) G_{3\tn{XX}} \nonumber \\
&+ B^2 \psi'
\left(
\frac{4 A^{\prime 2}}{A^2 r} - \frac{2 A''}{A r} - \frac{3 A' B'}{A B r}
- \frac{A'}{A r} - \frac{2 A' \psi''}{A \psi' r} - \frac{A'}{A B r^2}
\right)G_{4\tn{XX}} \nonumber \\
&+ \left(
\frac{2 A^{\prime 2} B}{A^2} - \frac{B A''}{A} - \frac{A' B'}{2 A}
- \frac{2 A' B}{A r} + \frac{2 B'}{r} + \frac{2\left(B - 1\right)}{r^2}
\right) G_{4\tn{X}\phi} \nonumber \\
&- B \psi' \left(
\frac{2 A'}{A} + \frac{B'}{B} + \frac{2\psi''}{\psi'} + \frac{4}{r}
\right) G_{4\tn{X}\phi\phi}
- B^2 \psi^{\prime 2}\left(
\frac{6 A'}{A r} +\frac{2 B'}{B r} - \frac{A' B'}{2 A B}
- \frac{A' \psi''}{A \psi'} +\frac{A^{\prime 2}}{2A^2}
+\frac{2}{r^2} + \frac{4 \psi''}{\psi' r}
\right) G_{4\tn{XX}\phi} \nonumber \\
&+ B^2 \psi^{\prime 3}\left(
\frac{A' B'}{Ar} + \frac{2 A' B \psi''}{A \psi' r} - \frac{A^{\prime 2} B}{A^2 r} - \frac{A' B}{A r^2}
\right) G_{4\tn{XXX}}\nonumber \\
&+ B^2 \left(
\frac{3 A^{\prime 2}}{4 A^2 r^2} - \frac{A''}{2 A r^2}
- \frac{3 A' B'}{4 A B r^2} - \frac{3 A^{\prime 2}}{4 A^2 B r^2}
+ \frac{A''}{2 A B r^2} + \frac{A' B'}{4 A B^2 r^2}
\right) G_{5\tn{X}} \nonumber \\
&+ B^2 \psi^{\prime 2}\left(
\frac{A'' B}{2 A r^2} - \frac{3 A^{\prime 2}B}{2 A^2 r^2}
+ \frac{3 A' B'}{2 A r^2} + \frac{3 A' B \psi''}{2 A \psi' r^2}
-\frac{A' B'}{4 A B r^2} - \frac{A' \psi''}{2 A \psi' r^2} +
\frac{A^{\prime 2}}{4 A^2 r^2}
\right) G_{5\tn{XX}} \nonumber \\
&+ B^2 \psi^{\prime} \left(
\frac{A''}{A r} - \frac{5 A^{\prime 2}}{2 A^2 r} + \frac{3 A' B'}{2 A B r}
- \frac{A'}{2 A r^2} - \frac{3 B'}{2 B r^2} + \frac{3 A'}{2 A B r^2}
+ \frac{A' \psi''}{A \psi' r} - \frac{\psi''}{\psi' r^2}
+ \frac{B'}{2 B^2 r^2} + \frac{\psi''}{\psi' B r^2}
\right) G_{5\tn{X}\phi} \nonumber \\
&- \left(
\frac{B-1}{r^2} + \frac{B'}{r}
\right) G_{5\phi \phi}
+ B^2 \phi^{\prime 2} \left(
\frac{2 A'}{A r} + \frac{B'}{B r} + \frac{2 \psi''}{\psi' r} + \frac{1}{r^2}
\right) G_{\tn{X}\phi\phi} \nonumber \\
&+ B^3 \psi^{\prime 3}\left(
\frac{3 A'}{2 A r^2} - \frac{A' B'}{2 A B r} - \frac{A' \psi''}{A \psi' r}
+ \frac{B'}{2 B r^2} + \frac{\psi''}{\psi' r^2} + \frac{A^{\prime 2}}{2 A^2 r}
\right) G_{5\tn{XX}\phi} \nonumber \\
&- B^4 \psi^{\prime 4}\left(
\frac{A' B'}{4 A B r^2} + \frac{A' \psi''}{2 A \psi' r^2} - \frac{A^{\prime 2}}{4 A^2 r^2}
\right) G_{5\tn{XXX}}\ .
\end{align}

\end{widetext}


\begin{thebibliography}{52}
\expandafter\ifx\csname natexlab\endcsname\relax\def\natexlab#1{#1}\fi
\expandafter\ifx\csname bibnamefont\endcsname\relax
  \def\bibnamefont#1{#1}\fi
\expandafter\ifx\csname bibfnamefont\endcsname\relax
  \def\bibfnamefont#1{#1}\fi
\expandafter\ifx\csname citenamefont\endcsname\relax
  \def\citenamefont#1{#1}\fi
\expandafter\ifx\csname url\endcsname\relax
  \def\url#1{\texttt{#1}}\fi
\expandafter\ifx\csname urlprefix\endcsname\relax\def\urlprefix{URL }\fi
\providecommand{\bibinfo}[2]{#2}
\providecommand{\eprint}[2][]{\url{#2}}

\bibitem[{\citenamefont{Will}(2014)}]{Will:2014kxa}
\bibinfo{author}{\bibfnamefont{C.~M.} \bibnamefont{Will}},
  \bibinfo{journal}{Living Rev. Rel.} \textbf{\bibinfo{volume}{17}},
  \bibinfo{pages}{4} (\bibinfo{year}{2014}), \eprint{1403.7377}.

\bibitem[{\citenamefont{Berti et~al.}(2015)}]{Berti:2015itd}
\bibinfo{author}{\bibfnamefont{E.}~\bibnamefont{Berti}} \bibnamefont{et~al.}
  (\bibinfo{year}{2015}), \eprint{1501.07274}.

\bibitem[{\citenamefont{Clifton et~al.}(2012)\citenamefont{Clifton, Ferreira,
  Padilla, and Skordis}}]{Clifton:2011jh}
\bibinfo{author}{\bibfnamefont{T.}~\bibnamefont{Clifton}},
  \bibinfo{author}{\bibfnamefont{P.~G.} \bibnamefont{Ferreira}},
  \bibinfo{author}{\bibfnamefont{A.}~\bibnamefont{Padilla}}, \bibnamefont{and}
  \bibinfo{author}{\bibfnamefont{C.}~\bibnamefont{Skordis}},
  \bibinfo{journal}{Phys. Rept.} \textbf{\bibinfo{volume}{513}},
  \bibinfo{pages}{1} (\bibinfo{year}{2012}), \eprint{1106.2476}.

\bibitem[{\citenamefont{Psaltis}(2008)}]{Psaltis:2008bb}
\bibinfo{author}{\bibfnamefont{D.}~\bibnamefont{Psaltis}},
  \bibinfo{journal}{Living Reviews in Relativity}  (\bibinfo{year}{2008}),
  \eprint{0806.1531}.

\bibitem[{\citenamefont{Yunes and Siemens}(2013)}]{Yunes:2013dva}
\bibinfo{author}{\bibfnamefont{N.}~\bibnamefont{Yunes}} \bibnamefont{and}
  \bibinfo{author}{\bibfnamefont{X.}~\bibnamefont{Siemens}},
  \bibinfo{journal}{Living Rev.Rel.} \textbf{\bibinfo{volume}{16}},
  \bibinfo{pages}{9} (\bibinfo{year}{2013}), \eprint{1304.3473}.

\bibitem[{\citenamefont{Gair et~al.}(2013)\citenamefont{Gair, Vallisneri,
  Larson, and Baker}}]{Gair:2012nm}
\bibinfo{author}{\bibfnamefont{J.~R.} \bibnamefont{Gair}},
  \bibinfo{author}{\bibfnamefont{M.}~\bibnamefont{Vallisneri}},
  \bibinfo{author}{\bibfnamefont{S.~L.} \bibnamefont{Larson}},
  \bibnamefont{and} \bibinfo{author}{\bibfnamefont{J.~G.} \bibnamefont{Baker}},
  \bibinfo{journal}{Living Rev.Rel.} \textbf{\bibinfo{volume}{16}},
  \bibinfo{pages}{7} (\bibinfo{year}{2013}), \eprint{1212.5575}.

\bibitem[{\citenamefont{Horndeski}(1974)}]{Horndeski:1974wa}
\bibinfo{author}{\bibfnamefont{G.~W.} \bibnamefont{Horndeski}},
  \bibinfo{journal}{Int.J.Theor.Phys.} \textbf{\bibinfo{volume}{10}},
  \bibinfo{pages}{363} (\bibinfo{year}{1974}).

\bibitem[{\citenamefont{Damour and Esposito-Farese}(1992)}]{Damour:1992we}
\bibinfo{author}{\bibfnamefont{T.}~\bibnamefont{Damour}} \bibnamefont{and}
  \bibinfo{author}{\bibfnamefont{G.}~\bibnamefont{Esposito-Farese}},
  \bibinfo{journal}{Class. Quant. Grav.} \textbf{\bibinfo{volume}{9}},
  \bibinfo{pages}{2093} (\bibinfo{year}{1992}).

\bibitem[{\citenamefont{Horbatsch et~al.}(2015)\citenamefont{Horbatsch, Silva,
  Gerosa, Pani, Berti, Gualtieri, and Sperhake}}]{Horbatsch:2015bua}
\bibinfo{author}{\bibfnamefont{M.}~\bibnamefont{Horbatsch}},
  \bibinfo{author}{\bibfnamefont{H.~O.} \bibnamefont{Silva}},
  \bibinfo{author}{\bibfnamefont{D.}~\bibnamefont{Gerosa}},
  \bibinfo{author}{\bibfnamefont{P.}~\bibnamefont{Pani}},
  \bibinfo{author}{\bibfnamefont{E.}~\bibnamefont{Berti}},
  \bibinfo{author}{\bibfnamefont{L.}~\bibnamefont{Gualtieri}},
  \bibnamefont{and} \bibinfo{author}{\bibfnamefont{U.}~\bibnamefont{Sperhake}}
  (\bibinfo{year}{2015}), \eprint{1505.07462}.

\bibitem[{\citenamefont{Padilla and Sivanesan}(2013)}]{Padilla:2012dx}
\bibinfo{author}{\bibfnamefont{A.}~\bibnamefont{Padilla}} \bibnamefont{and}
  \bibinfo{author}{\bibfnamefont{V.}~\bibnamefont{Sivanesan}},
  \bibinfo{journal}{JHEP} \textbf{\bibinfo{volume}{04}}, \bibinfo{pages}{032}
  (\bibinfo{year}{2013}), \eprint{1210.4026}.

\bibitem[{\citenamefont{Charmousis et~al.}(2014)\citenamefont{Charmousis,
  Kolyvaris, Papantonopoulos, and Tsoukalas}}]{Charmousis:2014zaa}
\bibinfo{author}{\bibfnamefont{C.}~\bibnamefont{Charmousis}},
  \bibinfo{author}{\bibfnamefont{T.}~\bibnamefont{Kolyvaris}},
  \bibinfo{author}{\bibfnamefont{E.}~\bibnamefont{Papantonopoulos}},
  \bibnamefont{and}
  \bibinfo{author}{\bibfnamefont{M.}~\bibnamefont{Tsoukalas}},
  \bibinfo{journal}{JHEP} \textbf{\bibinfo{volume}{07}}, \bibinfo{pages}{085}
  (\bibinfo{year}{2014}), \eprint{1404.1024}.

\bibitem[{\citenamefont{Nicolis et~al.}(2009)\citenamefont{Nicolis, Rattazzi,
  and Trincherini}}]{Nicolis:2008in}
\bibinfo{author}{\bibfnamefont{A.}~\bibnamefont{Nicolis}},
  \bibinfo{author}{\bibfnamefont{R.}~\bibnamefont{Rattazzi}}, \bibnamefont{and}
  \bibinfo{author}{\bibfnamefont{E.}~\bibnamefont{Trincherini}},
  \bibinfo{journal}{Phys. Rev.} \textbf{\bibinfo{volume}{D79}},
  \bibinfo{pages}{064036} (\bibinfo{year}{2009}), \eprint{0811.2197}.

\bibitem[{\citenamefont{Deffayet
  et~al.}(2009{\natexlab{a}})\citenamefont{Deffayet, Deser, and
  Esposito-Farese}}]{Deffayet:2009mn}
\bibinfo{author}{\bibfnamefont{C.}~\bibnamefont{Deffayet}},
  \bibinfo{author}{\bibfnamefont{S.}~\bibnamefont{Deser}}, \bibnamefont{and}
  \bibinfo{author}{\bibfnamefont{G.}~\bibnamefont{Esposito-Farese}},
  \bibinfo{journal}{Phys. Rev.} \textbf{\bibinfo{volume}{D80}},
  \bibinfo{pages}{064015} (\bibinfo{year}{2009}{\natexlab{a}}),
  \eprint{0906.1967}.

\bibitem[{\citenamefont{Kobayashi et~al.}(2011)\citenamefont{Kobayashi,
  Yamaguchi, and Yokoyama}}]{Kobayashi:2011nu}
\bibinfo{author}{\bibfnamefont{T.}~\bibnamefont{Kobayashi}},
  \bibinfo{author}{\bibfnamefont{M.}~\bibnamefont{Yamaguchi}},
  \bibnamefont{and} \bibinfo{author}{\bibfnamefont{J.}~\bibnamefont{Yokoyama}},
  \bibinfo{journal}{Prog. Theor. Phys.} \textbf{\bibinfo{volume}{126}},
  \bibinfo{pages}{511} (\bibinfo{year}{2011}), \eprint{1105.5723}.

\bibitem[{\citenamefont{Charmousis}(2015)}]{Charmousis:2014mia}
\bibinfo{author}{\bibfnamefont{C.}~\bibnamefont{Charmousis}},
  \bibinfo{journal}{Lect. Notes Phys.} \textbf{\bibinfo{volume}{892}},
  \bibinfo{pages}{25} (\bibinfo{year}{2015}), \eprint{1405.1612}.

\bibitem[{\citenamefont{Sushkov}(2009)}]{Sushkov:2009hk}
\bibinfo{author}{\bibfnamefont{S.~V.} \bibnamefont{Sushkov}},
  \bibinfo{journal}{Phys. Rev.} \textbf{\bibinfo{volume}{D80}},
  \bibinfo{pages}{103505} (\bibinfo{year}{2009}), \eprint{0910.0980}.

\bibitem[{\citenamefont{Saridakis and Sushkov}(2010)}]{Saridakis:2010mf}
\bibinfo{author}{\bibfnamefont{E.~N.} \bibnamefont{Saridakis}}
  \bibnamefont{and} \bibinfo{author}{\bibfnamefont{S.~V.}
  \bibnamefont{Sushkov}}, \bibinfo{journal}{Phys. Rev.}
  \textbf{\bibinfo{volume}{D81}}, \bibinfo{pages}{083510}
  (\bibinfo{year}{2010}), \eprint{1002.3478}.

\bibitem[{\citenamefont{Germani and Kehagias}(2010)}]{Germani:2010gm}
\bibinfo{author}{\bibfnamefont{C.}~\bibnamefont{Germani}} \bibnamefont{and}
  \bibinfo{author}{\bibfnamefont{A.}~\bibnamefont{Kehagias}},
  \bibinfo{journal}{Phys. Rev. Lett.} \textbf{\bibinfo{volume}{105}},
  \bibinfo{pages}{011302} (\bibinfo{year}{2010}), \eprint{1003.2635}.

\bibitem[{\citenamefont{Germani and Kehagias}(2011)}]{Germani:2010hd}
\bibinfo{author}{\bibfnamefont{C.}~\bibnamefont{Germani}} \bibnamefont{and}
  \bibinfo{author}{\bibfnamefont{A.}~\bibnamefont{Kehagias}},
  \bibinfo{journal}{Phys. Rev. Lett.} \textbf{\bibinfo{volume}{106}},
  \bibinfo{pages}{161302} (\bibinfo{year}{2011}), \eprint{1012.0853}.

\bibitem[{\citenamefont{Gubitosi and Linder}(2011)}]{Gubitosi:2011sg}
\bibinfo{author}{\bibfnamefont{G.}~\bibnamefont{Gubitosi}} \bibnamefont{and}
  \bibinfo{author}{\bibfnamefont{E.~V.} \bibnamefont{Linder}},
  \bibinfo{journal}{Phys. Lett.} \textbf{\bibinfo{volume}{B703}},
  \bibinfo{pages}{113} (\bibinfo{year}{2011}), \eprint{1106.2815}.

\bibitem[{\citenamefont{Kobayashi and Tanahashi}(2014)}]{Kobayashi:2014eva}
\bibinfo{author}{\bibfnamefont{T.}~\bibnamefont{Kobayashi}} \bibnamefont{and}
  \bibinfo{author}{\bibfnamefont{N.}~\bibnamefont{Tanahashi}},
  \bibinfo{journal}{PTEP} \textbf{\bibinfo{volume}{2014}},
  \bibinfo{pages}{073E02} (\bibinfo{year}{2014}), \eprint{1403.4364}.

\bibitem[{\citenamefont{Armendariz-Picon
  et~al.}(2001)\citenamefont{Armendariz-Picon, Mukhanov, and
  Steinhardt}}]{ArmendarizPicon:2000ah}
\bibinfo{author}{\bibfnamefont{C.}~\bibnamefont{Armendariz-Picon}},
  \bibinfo{author}{\bibfnamefont{V.~F.} \bibnamefont{Mukhanov}},
  \bibnamefont{and} \bibinfo{author}{\bibfnamefont{P.~J.}
  \bibnamefont{Steinhardt}}, \bibinfo{journal}{Phys. Rev.}
  \textbf{\bibinfo{volume}{D63}}, \bibinfo{pages}{103510}
  (\bibinfo{year}{2001}), \eprint{astro-ph/0006373}.

\bibitem[{\citenamefont{Armendariz-Picon
  et~al.}(1999)\citenamefont{Armendariz-Picon, Damour, and
  Mukhanov}}]{ArmendarizPicon:1999rj}
\bibinfo{author}{\bibfnamefont{C.}~\bibnamefont{Armendariz-Picon}},
  \bibinfo{author}{\bibfnamefont{T.}~\bibnamefont{Damour}}, \bibnamefont{and}
  \bibinfo{author}{\bibfnamefont{V.~F.} \bibnamefont{Mukhanov}},
  \bibinfo{journal}{Phys. Lett.} \textbf{\bibinfo{volume}{B458}},
  \bibinfo{pages}{209} (\bibinfo{year}{1999}), \eprint{hep-th/9904075}.

\bibitem[{\citenamefont{Alishahiha et~al.}(2004)\citenamefont{Alishahiha,
  Silverstein, and Tong}}]{Alishahiha:2004eh}
\bibinfo{author}{\bibfnamefont{M.}~\bibnamefont{Alishahiha}},
  \bibinfo{author}{\bibfnamefont{E.}~\bibnamefont{Silverstein}},
  \bibnamefont{and} \bibinfo{author}{\bibfnamefont{D.}~\bibnamefont{Tong}},
  \bibinfo{journal}{Phys. Rev.} \textbf{\bibinfo{volume}{D70}},
  \bibinfo{pages}{123505} (\bibinfo{year}{2004}), \eprint{hep-th/0404084}.

\bibitem[{\citenamefont{Deffayet
  et~al.}(2009{\natexlab{b}})\citenamefont{Deffayet, Esposito-Farese, and
  Vikman}}]{Deffayet:2009wt}
\bibinfo{author}{\bibfnamefont{C.}~\bibnamefont{Deffayet}},
  \bibinfo{author}{\bibfnamefont{G.}~\bibnamefont{Esposito-Farese}},
  \bibnamefont{and} \bibinfo{author}{\bibfnamefont{A.}~\bibnamefont{Vikman}},
  \bibinfo{journal}{Phys. Rev.} \textbf{\bibinfo{volume}{D79}},
  \bibinfo{pages}{084003} (\bibinfo{year}{2009}{\natexlab{b}}),
  \eprint{0901.1314}.

\bibitem[{\citenamefont{Rinaldi}(2012)}]{Rinaldi:2012vy}
\bibinfo{author}{\bibfnamefont{M.}~\bibnamefont{Rinaldi}},
  \bibinfo{journal}{Phys. Rev.} \textbf{\bibinfo{volume}{D86}},
  \bibinfo{pages}{084048} (\bibinfo{year}{2012}), \eprint{1208.0103}.

\bibitem[{\citenamefont{Minamitsuji}(2014{\natexlab{a}})}]{Minamitsuji:2013ura}
\bibinfo{author}{\bibfnamefont{M.}~\bibnamefont{Minamitsuji}},
  \bibinfo{journal}{Phys. Rev.} \textbf{\bibinfo{volume}{D89}},
  \bibinfo{pages}{064017} (\bibinfo{year}{2014}{\natexlab{a}}),
  \eprint{1312.3759}.

\bibitem[{\citenamefont{Anabalon et~al.}(2014)\citenamefont{Anabalon, Cisterna,
  and Oliva}}]{Anabalon:2013oea}
\bibinfo{author}{\bibfnamefont{A.}~\bibnamefont{Anabalon}},
  \bibinfo{author}{\bibfnamefont{A.}~\bibnamefont{Cisterna}}, \bibnamefont{and}
  \bibinfo{author}{\bibfnamefont{J.}~\bibnamefont{Oliva}},
  \bibinfo{journal}{Phys. Rev.} \textbf{\bibinfo{volume}{D89}},
  \bibinfo{pages}{084050} (\bibinfo{year}{2014}), \eprint{1312.3597}.

\bibitem[{\citenamefont{Hui and Nicolis}(2013)}]{Hui:2012qt}
\bibinfo{author}{\bibfnamefont{L.}~\bibnamefont{Hui}} \bibnamefont{and}
  \bibinfo{author}{\bibfnamefont{A.}~\bibnamefont{Nicolis}},
  \bibinfo{journal}{Phys. Rev. Lett.} \textbf{\bibinfo{volume}{110}},
  \bibinfo{pages}{241104} (\bibinfo{year}{2013}), \eprint{1202.1296}.

\bibitem[{\citenamefont{Sotiriou and
  Zhou}(2014{\natexlab{a}})}]{Sotiriou:2013qea}
\bibinfo{author}{\bibfnamefont{T.~P.} \bibnamefont{Sotiriou}} \bibnamefont{and}
  \bibinfo{author}{\bibfnamefont{S.-Y.} \bibnamefont{Zhou}},
  \bibinfo{journal}{Phys. Rev. Lett.} \textbf{\bibinfo{volume}{112}},
  \bibinfo{pages}{251102} (\bibinfo{year}{2014}{\natexlab{a}}),
  \eprint{1312.3622}.

\bibitem[{\citenamefont{Sotiriou and
  Zhou}(2014{\natexlab{b}})}]{Sotiriou:2014pfa}
\bibinfo{author}{\bibfnamefont{T.~P.} \bibnamefont{Sotiriou}} \bibnamefont{and}
  \bibinfo{author}{\bibfnamefont{S.-Y.} \bibnamefont{Zhou}},
  \bibinfo{journal}{Phys. Rev.} \textbf{\bibinfo{volume}{D90}},
  \bibinfo{pages}{124063} (\bibinfo{year}{2014}{\natexlab{b}}),
  \eprint{1408.1698}.

\bibitem[{\citenamefont{Kanti et~al.}(1996)\citenamefont{Kanti, Mavromatos,
  Rizos, Tamvakis, and Winstanley}}]{Kanti:1995vq}
\bibinfo{author}{\bibfnamefont{P.}~\bibnamefont{Kanti}},
  \bibinfo{author}{\bibfnamefont{N.}~\bibnamefont{Mavromatos}},
  \bibinfo{author}{\bibfnamefont{J.}~\bibnamefont{Rizos}},
  \bibinfo{author}{\bibfnamefont{K.}~\bibnamefont{Tamvakis}}, \bibnamefont{and}
  \bibinfo{author}{\bibfnamefont{E.}~\bibnamefont{Winstanley}},
  \bibinfo{journal}{Phys.Rev.} \textbf{\bibinfo{volume}{D54}},
  \bibinfo{pages}{5049} (\bibinfo{year}{1996}), \eprint{hep-th/9511071}.

\bibitem[{\citenamefont{Pani and Cardoso}(2009)}]{Pani:2009wy}
\bibinfo{author}{\bibfnamefont{P.}~\bibnamefont{Pani}} \bibnamefont{and}
  \bibinfo{author}{\bibfnamefont{V.}~\bibnamefont{Cardoso}},
  \bibinfo{journal}{Phys.Rev.} \textbf{\bibinfo{volume}{D79}},
  \bibinfo{pages}{084031} (\bibinfo{year}{2009}), \eprint{0902.1569}.

\bibitem[{\citenamefont{Ayzenberg and Yunes}(2014)}]{Ayzenberg:2014aka}
\bibinfo{author}{\bibfnamefont{D.}~\bibnamefont{Ayzenberg}} \bibnamefont{and}
  \bibinfo{author}{\bibfnamefont{N.}~\bibnamefont{Yunes}},
  \bibinfo{journal}{Phys. Rev.} \textbf{\bibinfo{volume}{D90}},
  \bibinfo{pages}{044066} (\bibinfo{year}{2014}), \bibinfo{note}{[Erratum:
  Phys. Rev.D91,no.6,069905(2015)]}, \eprint{1405.2133}.

\bibitem[{\citenamefont{Maselli et~al.}(2015)\citenamefont{Maselli, Pani,
  Gualtieri, and Ferrari}}]{Maselli:2015tta}
\bibinfo{author}{\bibfnamefont{A.}~\bibnamefont{Maselli}},
  \bibinfo{author}{\bibfnamefont{P.}~\bibnamefont{Pani}},
  \bibinfo{author}{\bibfnamefont{L.}~\bibnamefont{Gualtieri}},
  \bibnamefont{and} \bibinfo{author}{\bibfnamefont{V.}~\bibnamefont{Ferrari}}
  (\bibinfo{year}{2015}), \eprint{1507.00680}.

\bibitem[{\citenamefont{Kleihaus et~al.}(2011)\citenamefont{Kleihaus, Kunz, and
  Radu}}]{Kleihaus:2011tg}
\bibinfo{author}{\bibfnamefont{B.}~\bibnamefont{Kleihaus}},
  \bibinfo{author}{\bibfnamefont{J.}~\bibnamefont{Kunz}}, \bibnamefont{and}
  \bibinfo{author}{\bibfnamefont{E.}~\bibnamefont{Radu}},
  \bibinfo{journal}{Phys. Rev. Lett.} \textbf{\bibinfo{volume}{106}},
  \bibinfo{pages}{151104} (\bibinfo{year}{2011}), \eprint{1101.2868}.

\bibitem[{\citenamefont{Kleihaus et~al.}(2014)\citenamefont{Kleihaus, Kunz, and
  Mojica}}]{Kleihaus:2014lba}
\bibinfo{author}{\bibfnamefont{B.}~\bibnamefont{Kleihaus}},
  \bibinfo{author}{\bibfnamefont{J.}~\bibnamefont{Kunz}}, \bibnamefont{and}
  \bibinfo{author}{\bibfnamefont{S.}~\bibnamefont{Mojica}},
  \bibinfo{journal}{Phys. Rev.} \textbf{\bibinfo{volume}{D90}},
  \bibinfo{pages}{061501} (\bibinfo{year}{2014}), \eprint{1407.6884}.

\bibitem[{\citenamefont{Babichev and Charmousis}(2014)}]{Babichev:2013cya}
\bibinfo{author}{\bibfnamefont{E.}~\bibnamefont{Babichev}} \bibnamefont{and}
  \bibinfo{author}{\bibfnamefont{C.}~\bibnamefont{Charmousis}},
  \bibinfo{journal}{JHEP} \textbf{\bibinfo{volume}{1408}}, \bibinfo{pages}{106}
  (\bibinfo{year}{2014}), \eprint{1312.3204}.

\bibitem[{\citenamefont{Babichev et~al.}(2015)\citenamefont{Babichev,
  Charmousis, and Hassaine}}]{Babichev:2015rva}
\bibinfo{author}{\bibfnamefont{E.}~\bibnamefont{Babichev}},
  \bibinfo{author}{\bibfnamefont{C.}~\bibnamefont{Charmousis}},
  \bibnamefont{and} \bibinfo{author}{\bibfnamefont{M.}~\bibnamefont{Hassaine}},
  \bibinfo{journal}{JCAP} \textbf{\bibinfo{volume}{1505}}, \bibinfo{pages}{031}
  (\bibinfo{year}{2015}), \eprint{1503.02545}.

\bibitem[{\citenamefont{Kolyvaris et~al.}(2012)\citenamefont{Kolyvaris,
  Koutsoumbas, Papantonopoulos, and Siopsis}}]{Kolyvaris:2011fk}
\bibinfo{author}{\bibfnamefont{T.}~\bibnamefont{Kolyvaris}},
  \bibinfo{author}{\bibfnamefont{G.}~\bibnamefont{Koutsoumbas}},
  \bibinfo{author}{\bibfnamefont{E.}~\bibnamefont{Papantonopoulos}},
  \bibnamefont{and} \bibinfo{author}{\bibfnamefont{G.}~\bibnamefont{Siopsis}},
  \bibinfo{journal}{Class. Quant. Grav.} \textbf{\bibinfo{volume}{29}},
  \bibinfo{pages}{205011} (\bibinfo{year}{2012}), \eprint{1111.0263}.

\bibitem[{\citenamefont{Hartle}(1967)}]{Hartle:1967he}
\bibinfo{author}{\bibfnamefont{J.~B.} \bibnamefont{Hartle}},
  \bibinfo{journal}{Astrophys.J.} \textbf{\bibinfo{volume}{150}},
  \bibinfo{pages}{1005} (\bibinfo{year}{1967}).

\bibitem[{\citenamefont{Hartle and Thorne}(1968)}]{Hartle:1968si}
\bibinfo{author}{\bibfnamefont{J.~B.} \bibnamefont{Hartle}} \bibnamefont{and}
  \bibinfo{author}{\bibfnamefont{K.~S.} \bibnamefont{Thorne}},
  \bibinfo{journal}{Astrophys.J.} \textbf{\bibinfo{volume}{153}},
  \bibinfo{pages}{807} (\bibinfo{year}{1968}).

\bibitem[{\citenamefont{Kobayashi et~al.}(2012)\citenamefont{Kobayashi,
  Motohashi, and Suyama}}]{Kobayashi:2012kh}
\bibinfo{author}{\bibfnamefont{T.}~\bibnamefont{Kobayashi}},
  \bibinfo{author}{\bibfnamefont{H.}~\bibnamefont{Motohashi}},
  \bibnamefont{and} \bibinfo{author}{\bibfnamefont{T.}~\bibnamefont{Suyama}},
  \bibinfo{journal}{Phys. Rev.} \textbf{\bibinfo{volume}{D85}},
  \bibinfo{pages}{084025} (\bibinfo{year}{2012}), \eprint{1202.4893}.

\bibitem[{\citenamefont{Kobayashi et~al.}(2014)\citenamefont{Kobayashi,
  Motohashi, and Suyama}}]{Kobayashi:2014wsa}
\bibinfo{author}{\bibfnamefont{T.}~\bibnamefont{Kobayashi}},
  \bibinfo{author}{\bibfnamefont{H.}~\bibnamefont{Motohashi}},
  \bibnamefont{and} \bibinfo{author}{\bibfnamefont{T.}~\bibnamefont{Suyama}},
  \bibinfo{journal}{Phys. Rev.} \textbf{\bibinfo{volume}{D89}},
  \bibinfo{pages}{084042} (\bibinfo{year}{2014}), \eprint{1402.6740}.

\bibitem[{\citenamefont{{Maselli} et~al.}(2015)}]{MaselliPrep}
\bibinfo{author}{\bibfnamefont{A.}~\bibnamefont{{Maselli}}}
  \bibnamefont{et~al.} (\bibinfo{year}{2015}), \bibinfo{note}{in preparation}.

\bibitem[{\citenamefont{Berti et~al.}(2009)\citenamefont{Berti, Cardoso, and
  Starinets}}]{Berti:2009kk}
\bibinfo{author}{\bibfnamefont{E.}~\bibnamefont{Berti}},
  \bibinfo{author}{\bibfnamefont{V.}~\bibnamefont{Cardoso}}, \bibnamefont{and}
  \bibinfo{author}{\bibfnamefont{A.~O.} \bibnamefont{Starinets}},
  \bibinfo{journal}{Class. Quant. Grav.} \textbf{\bibinfo{volume}{26}},
  \bibinfo{pages}{163001} (\bibinfo{year}{2009}), \eprint{0905.2975}.

\bibitem[{\citenamefont{Minamitsuji}(2014{\natexlab{b}})}]{Minamitsuji:2014hha}
\bibinfo{author}{\bibfnamefont{M.}~\bibnamefont{Minamitsuji}},
  \bibinfo{journal}{Gen. Rel. Grav.} \textbf{\bibinfo{volume}{46}},
  \bibinfo{pages}{1785} (\bibinfo{year}{2014}{\natexlab{b}}),
  \eprint{1407.4901}.

\bibitem[{\citenamefont{Pani}(2013)}]{Pani:2013pma}
\bibinfo{author}{\bibfnamefont{P.}~\bibnamefont{Pani}},
  \bibinfo{journal}{Int.J.Mod.Phys.} \textbf{\bibinfo{volume}{A28}},
  \bibinfo{pages}{1340018} (\bibinfo{year}{2013}), \eprint{1305.6759}.

\bibitem[{\citenamefont{Pani et~al.}(2011)\citenamefont{Pani, Berti, Cardoso,
  and Read}}]{Pani:2011xm}
\bibinfo{author}{\bibfnamefont{P.}~\bibnamefont{Pani}},
  \bibinfo{author}{\bibfnamefont{E.}~\bibnamefont{Berti}},
  \bibinfo{author}{\bibfnamefont{V.}~\bibnamefont{Cardoso}}, \bibnamefont{and}
  \bibinfo{author}{\bibfnamefont{J.}~\bibnamefont{Read}},
  \bibinfo{journal}{Phys. Rev.} \textbf{\bibinfo{volume}{D84}},
  \bibinfo{pages}{104035} (\bibinfo{year}{2011}), \eprint{1109.0928}.

\bibitem[{\citenamefont{Cisterna et~al.}(2015)\citenamefont{Cisterna, Delsate,
  and Rinaldi}}]{Cisterna:2015yla}
\bibinfo{author}{\bibfnamefont{A.}~\bibnamefont{Cisterna}},
  \bibinfo{author}{\bibfnamefont{T.}~\bibnamefont{Delsate}}, \bibnamefont{and}
  \bibinfo{author}{\bibfnamefont{M.}~\bibnamefont{Rinaldi}}
  (\bibinfo{year}{2015}), \eprint{1504.05189}.

\bibitem[{\citenamefont{Damour and Esposito-Farese}(1993)}]{Damour:1993hw}
\bibinfo{author}{\bibfnamefont{T.}~\bibnamefont{Damour}} \bibnamefont{and}
  \bibinfo{author}{\bibfnamefont{G.}~\bibnamefont{Esposito-Farese}},
  \bibinfo{journal}{Phys. Rev. Lett.} \textbf{\bibinfo{volume}{70}},
  \bibinfo{pages}{2220} (\bibinfo{year}{1993}).

\bibitem[{\citenamefont{Chen et~al.}(2015)\citenamefont{Chen, Suyama, and
  Yokoyama}}]{Chen:2015zmx}
\bibinfo{author}{\bibfnamefont{P.}~\bibnamefont{Chen}},
  \bibinfo{author}{\bibfnamefont{T.}~\bibnamefont{Suyama}}, \bibnamefont{and}
  \bibinfo{author}{\bibfnamefont{J.}~\bibnamefont{Yokoyama}}
  (\bibinfo{year}{2015}), \eprint{1508.01384}.

\end{thebibliography}

\end{document}